\documentclass[10pt]{article}

\usepackage{xspace}
\usepackage{ragged2e}
\usepackage{url}
\usepackage{mathtools}
\usepackage{amssymb}
\usepackage{amsthm}
\usepackage{empheq}
\usepackage{latexsym}
\usepackage{enumitem}
\usepackage{eurosym}
\usepackage{dsfont}
\usepackage{appendix}
\usepackage{color} 
\usepackage[unicode]{hyperref}
\usepackage{frcursive}
\usepackage[utf8]{inputenc}
\usepackage[T1]{fontenc}
\usepackage{geometry}
\usepackage{multirow}
\usepackage{todonotes}
\usepackage{lmodern}
\usepackage{subcaption}
\usepackage{anyfontsize}
\usepackage{stmaryrd}
\usepackage{natbib}
\usepackage{cleveref}
\usepackage[english]{babel}
\usepackage[english=british]{csquotes}
\usepackage{float}

\everymath{\displaystyle}

\setcitestyle{numbers,open={[},close={]}}

\definecolor{red}{rgb}{0.7,0.15,0.15}
\definecolor{green}{rgb}{0,0.5,0}
\definecolor{blue}{rgb}{0,0,0.7}
\hypersetup{colorlinks, linkcolor={red},citecolor={green}, urlcolor={blue}}
			
\makeatletter \@addtoreset{equation}{section}

\newtheorem{theorem}{Theorem}[section]

\newtheorem{corollary}[theorem]{Corollary}

\newtheorem{definition}[theorem]{Definition}
\newtheorem{remark}[theorem]{Remark}

\def \F{\mathbb{F}}
\def \G{\mathbb{G}}

\def \L{\mathbb{L}}

\def \N{\mathbb{N}}

\def \P{\mathbb{P}}

\def \R{\mathbb{R}}

\def \Z{\mathbb{Z}}

\def\Ac{{\cal A}}

\def\Dc{{\cal D}}

\def\Gc{{\cal G}}

\def\Kc{{\cal K}}
\def\Lc{{\cal L}}
\def\Mc{{\cal M}}

\def\Pc{{\cal P}}

\def\Vc{{\cal V}}
\def\Wc{{\cal W}}

\def\d{\mathrm{d}}

\title{A mean-field game of market-making against strategic traders\footnote{This work benefits from the financial support of the Chaire Deep Finance and Statistics. The authors would like to thank Idris Kharroubi, Gilles Pagès and Huyên Pham for insightful discussions on the subject.}}
\author{Bastien {\sc Baldacci}\footnote{Quantitative Advisory Solutions, bastien.baldacci.qas@protonmail.com.}\and 
Philippe {\sc Bergault}\footnote{\'Ecole Polytechnique, CMAP, Route de Saclay, 91128, Palaiseau Cedex, France, philippe.bergault@polytechnique.edu.}\and Dylan {\sc Possamaï}\footnote{ETH Z\"urich, Department of Mathematics, R\"amistrasse 101, 8092 Z\"urich, Switzerland, dylan.possamai@math.ethz.ch.} }

\begin{document}

\maketitle

\begin{abstract}
We design a market-making model à la \citeauthor*{avellaneda2008high} in which the market-takers act strategically, in the sense that they design their trading strategy based on an exogenous trading signal. The market-maker chooses her quotes based on the average market-takers' behaviour, modelled through a mean-field interaction. We derive, up to the resolution of a coupled HJB--Fokker--Planck system, the optimal controls of the market-maker and the representative market-taker. This approach is flexible enough to incorporate different behaviours for the market-takers and takes into account the impact of their strategies on the price process. 

 \vspace{5mm}
\end{abstract}

\section{Introduction}
On financial markets, a market-maker is usually defined as a liquidity provider. In a nutshell, it is a market participant who provides bid and ask (\emph{i.e.} buy and sell) prices on one or several assets. The market-maker makes profit by earning the price difference between her buy and sell orders, called the bid--ask spread. She faces the following optimisation problem: being rarely executed at a very high spread, or being often executed at a very low spread. In the first case, the marked-to-market gain will be high, and conversely in the second case. On most of the order-driven markets, such as equities, the market-making activity is provided by a small number of high-frequency trading firms which are the counterpart of the vast majority of the transactions, see for example \citeauthor*{megarbane2017behavior} \cite{megarbane2017behavior}. They quote almost continuously bid and ask prices at the best limits, and their strategies are mostly based on either their superior proprietary technology, or on arbitrage strategies on cross-listed securities based on microstructure signals, see for instance \citeauthor*{baldacci2020adaptive} \cite{baldacci2020adaptive}. On quote-driven markets, such as for many fixed income products and single stock options, the liquidity-takers, who have a fair price in mind, first request a price for a given transaction size of the asset to a market-maker, often called dealer. The dealer answers by providing a price for the corresponding size, which is then accepted or rejected by the client. It is common that, similarly to order-driven markets, a small number of dealers provide the vast majority of liquidity to a large set of clients. On some specific products, such as options on single stocks, there can even be a monopolistic situation where there is a single liquidity provider\footnote{For example, Citadel Securities operates as a single market-maker for over 4,000 U.S. listed-options names, representing 99\% of traded volume, see \url{ https://www.citadelsecurities.com/products/equities-and-options/}.}.

\medskip
The optimal market-making problem has been the subject of a vast academic literature. Seminal references are \citeauthor*{grossman1988liquidity} \cite{grossman1988liquidity}, and \citeauthor*{ho1981optimal} \cite{ho1981optimal}. In particular, the authors of \cite{ho1981optimal} studied the behaviour of a market-maker confronting a stochastic demand, using stochastic control theory. This framework has inspired the well-known model of \citeauthor*{avellaneda2008high} \cite{avellaneda2008high}, which was designed to be applicable to order-driven market at the high-frequency scale. However, due to the continuous nature of the market-maker's spreads, and the assumption that the underlying asset is a diffusion process, this model turned out to also be suitable for quote-driven markets, such as corporate bonds markets. By providing a rigorous analysis of the stochastic control problem of \cite{avellaneda2008high}, \citeauthor*{gueant2013dealing} \cite{gueant2013dealing} showed, in the case of a CARA utility function, that the market-maker's problem boiled down to a system of linear ordinary differential equations. More recent contributions to the market-making literature can be found in the works of \citeauthor*{cartea2015algorithmic} \cite{cartea2015algorithmic}, \citeauthor*{cartea2016incorporating} \cite{cartea2016incorporating}, or \citeauthor*{cartea2017algorithmic} \cite{cartea2017algorithmic,cartea2018enhancing}, who enriched the initial model by introducing $\alpha$ signals, ambiguity aversion, competition with other agents... In all these articles, the authors consider a risk-adjusted expectation instead of a CARA utility function, which leads to the same optimal quotes with a suitable transformation of the intensity functions, see \citeauthor*{manziuk2019optimal} \cite{manziuk2019optimal}. More recently, multi-asset market-making has been addressed through reinforcement learning, see \citeauthor*{gueant2019deep} \cite{gueant2019deep}, and dimensionality reduction techniques, see \citeauthor*{bergault2021size} \cite{bergault2021size}, \citeauthor*{bergault2021closed} \cite{bergault2021closed}, or \citeauthor*{baldacci2021algorithmic} \cite{baldacci2021algorithmic} for the special case of option market-making. All these models are well-suited for OTC markets and for order-driven markets in the case of small tick assets, but are not straightforwardly applicable to design market-making strategies on order books with a large tick size, where the control process lies in a discrete tick grid. One of the main contributions to this literature is \citeauthor*{guilbaud2013optimal} \cite{guilbaud2013optimal}, where the authors build a model where a market-maker can send both limit and market orders on the order book of a single asset.\footnote{A similar behaviour in the case of OTC market-making has been studied in \citeauthor*{barzykin2021algorithmic} \cite{barzykin2021algorithmic}, in which the agent can choose to trade `actively' without waiting for a client, see also \citeauthor*{barzykin2021market} \cite{barzykin2021market}.  } In a similar vein, the model with uncertainty zones introduced by \citeauthor*{robert2011new} \cite{robert2011new} to reproduce accurately the high-frequency behaviour of an asset, has been incorporated into a market-making model, see \citeauthor*{baldacci2020bid} \cite{baldacci2020bid}. 

\medskip
A striking feature of all the aforementioned models is that the market-takers are systematically assumed to be passive, in the sense that they only respond to a quote proposed by the market-maker through the intensity of a point process. The case of market-making with competitive agents has rarely been addressed in the literature.\footnote{Notice nonetheless that a framework for multi market-makers competing with each other to propose the best price to the clients is developed in \citeauthor*{baldacci2021optimal} \cite{baldacci2021optimal} in the context of make-take fees policy.} The reason is that accounting for multiple agents leads to high dimensional problems with complicated interactions, and typically renders the analysis intractable using standard numerical methods in stochastic optimal control. In the context of optimal trading, interactions between a large number of agents have been modelled through the use of mean-field games. This is one possible solution to model more complex market interactions, because of its tractability compared to multi-agent models. The classical optimal execution problem à la \citeauthor*{almgren2001optimal} \cite{almgren2001optimal} is thus addressed by controlling McKean--Vlasov equations in \citeauthor*{cardaliaguet2018mean} \cite{cardaliaguet2018mean}: the trader faces uncertainty with respect to price changes because of his actions but also has to deal with price changes due to other similar market participants, impacting the prices permanently and acting strategically. Therefore, the distribution of the controls appears inside the drift of the asset's price process. In \citeauthor*{huang2019mean} \cite{huang2019mean}, the authors extend the previous work on optimal trading using a major--minor mean-field game framework, where a major agent is liquidating a large portion of shares, and a large number of minor agents trade along with the major agent. Mean-field games have also been applied by \citeauthor*{casgrain2018mean} \cite{casgrain2018mean,casgrain2020mean} to the case of heterogeneous traders who aim at performing optimal execution. All these models have in common that the mean-field interaction is through the drift of the asset's price process. Moreover, in the case of major--minor optimal trading in \citeauthor*{huang2019mean} \cite{huang2019mean}, both the major and the minor agents are traders who only differ by their influence on the market. As stated previously, there is no strategic market-takers in the extensions of \citeauthor{avellaneda2008high} \cite{avellaneda2008high} due to the difficulties in modelling the interactions between the market-maker and several market-takers. Having a very granular model will lead to a high-dimensional stochastic control problems, intractable with standard numerical methods. Therefore, we propose in this paper to employ a major--minor mean-field game where the market-maker plays the role of the major player, whereas the strategic market-takers are modelled through a mean-field of minor agents.  

\medskip
We emphasise that many recent papers also make use of the theory of mean-field games with a major player, introduced and studied deeply by \citeauthor*{buckdahn2014nonlinear} \cite{buckdahn2014nonlinear}, \citeauthor*{cardaliaguet2020remarks} \cite{cardaliaguet2020remarks}, \citeauthor*{carmona2016finite} \cite{carmona2016finite,carmona2017alternative}, \citeauthor*{carmona2016probabilistic} \cite{carmona2016probabilistic}, and  \citeauthor*{lasry2018mean} \cite{lasry2018mean}. As in standard mean-field games and optimal control theory, linear--quadratic problems are of particular interest in the presence of a major player facing a mean-field of minor agents as they generally boil down to a system of ODEs of the Riccati type, that can sometimes be solved explicitly. Those problems have been studied in particular in \citeauthor*{huang2010large} \cite{huang2010large, huang2021linear}, \citeauthor*{huang2016backward} \cite{huang2016backward}, \citeauthor*{huang2019linear} \cite{huang2019linear}, but also in \citeauthor*{firoozi2021epsilon} \cite{firoozi2021epsilon, firoozi2019belief} and \citeauthor*{firoozi2020convex} \cite{firoozi2020convex}.

\medskip
Coming back to the present paper, our main goal is to address the problem of a market-maker providing the vast majority of the liquidity on an underlying asset. The market-maker faces a very high number of traders, or market-takers, with possibly different behaviours such as momentum or mean reverting strategies. Each market-taker acts strategically in the sense that they send request to the market-maker with a fair price in mind, which results from their own optimisation, depending on their inventory and their trading signal. The market-maker proposes quotes based on her inventory process, and the behaviour of her counterparts, which are aggregated through a mean-field game. The problem thus takes the form of a mean-field game with controlled jumps: the intensities of the order arrivals of all market participants are averaged through a single point process, and the market-maker increases her execution rate if her quote is close to the fair price resulting from the mean-field equilibrium of the minor players. Based on the mean-field equilibrium between the traders, the market-maker computes the average fair price of the mean-field. If the fair bid price of the minor players is higher than her ask quote, for instance, the intensity of arrival orders at the ask for the market-maker will decrease and conversely. We characterise completely the problem by a system of Fokker--Planck equations for the mean-field game associated to the minor players, and a (decoupled) HJB equation for the major player, which can be efficiently solved via classic numerical methods on grids. Up to the resolution of these PDEs, we derive the optimal strategies of both the representative market-taker, and the market-maker. To the best of our knowledge, this is the first work on optimal market-making taking into account the strategic behaviour of market-takers. We believe that the use of mean-field games to represent the interaction between market-maker and market-takers can lead to substantial improvement of the usual optimal market-making models, without increasing drastically the numerical complexity. The numerical results with a set of reasonable market parameters exhibit some interesting features. The probability distribution of the market-takers' inventory shifts with the exogenous trading signal and the market-maker's strategy is adjusted with respect to this flow. Even with an empty inventory, the market-takers prefer to have a non-zero signal as it will allow them to make more profit over the trading horizon: the worst-case for a market-taker is a flat signal. Moreover, when market-takers have a very volatile signal, the distribution of inventories remains symmetric around $0$ but with two `bumps' representing their incapacity to react rapidly to a change of signal. These results enable to enhance the seminal work of \citeauthor*{avellaneda2008high} \cite{avellaneda2008high} in the sense that the behaviour of market-takers is not fixed as a function of the quotes of the market-maker only, but as a function of their own view on the price process as well as their current inventory. This is of particular importance as it leads to a more accurate estimation of the PnL of the strategy of the market-maker. Moreover, our model also provides the optimal strategy of an informed market-taker, which is absent in \citeauthor*{avellaneda2008high} \cite{avellaneda2008high}. 

\medskip
The article is constructed as follows: in \Cref{sec_N_players}, we introduce informally an $N$-player version of the model, which justifies the presentation in \Cref{mfg_proba_framework} of the mean-field probabilistic framework. In \Cref{MarkovCasesec}, we solve those problems in the case of Markovian controls, while \Cref{numsec} is devoted to the numerical results.

\medskip
{\bf Notations:} \noindent
Let $\N^\star$ be the set of all positive integers. For any $(\ell,c)\in \mathbb{N}^{\star} \times \mathbb{N}^{\star}$, $\mathcal{M}_{\ell,c}(\mathbb{R})$ will denote the space of $\ell\times c$ matrices with real entries. Elements of the matrix $M\in \mathcal{M}_{\ell,c}(\R)$ are denoted by $(M^{i,j})_{(i,j)\in\{1,\dots,\ell\}\times\{1,\dots,c\}}$ and the transpose of $M$ is denoted by $M^{\top}$. We identify $\mathcal{M}_{\ell,1}$ with $\mathbb{R}^{\ell}$. When $\ell=c$, we let $\mathcal{M}_{\ell}(\mathbb{R}):=\mathcal{M}_{\ell,\ell}(\mathbb{R})$. For any $x\in \mathcal{M}_{\ell,c}(\mathbb{R})$, and for any $j\in\{1,\dots,c\}$, $x^{j}\in \mathbb{R}^{\ell}$ is the $j$-th column of $M$. Moreover, for any $x \in \mathcal{M}_{\ell,c}(\mathbb{R})$ and any $j\in\{1,\dots,c\}$, $x^{-j}\in \mathcal{M}_{\ell,c-1}(\mathbb{R})$ denotes the matrix $x$ without the $j$-th column. For a pair $(x^0,x) \in \mathbb R^\ell \times \mathcal M_{\ell, c}(\mathbb{R})$, we denote by $x^{-0}$ the matrix $x^{-0}:= x$. For any $x\in \mathcal{M}_{\ell,c}(\mathbb{R})$ and $y\in \mathbb{R}^{\ell}$, we also define for $i\in\{1,\dots,\ell\}$, $y\otimes_{i}x \in \mathcal{M}_{\ell,c+1}(\mathbb{R})$ as the matrix whose first $i-1$ columns are equal to the first $i-1$ columns of $x$, such that for $j\in\{i+1,\dots,c+1\}$, its $j$-th column is equal to the $(j-1)$-th column of $x$, and whose $i$-th column is equal to $y$. We extend naturally this definition to $\mathcal{M}_{\ell,c}(\mathbb{R})$-valued processes. For $(a,b) \in \mathbb R^2$ with $a\leq b$, we denote $\llbracket a,b \rrbracket := [a,b] \cap \mathbb Z$. For a generic finite-dimensional Euclidean space $E$, a generic filtered probability space $(\Ac,\Gc,\G,\Pi)$, we let $\L^0(E,\G)$ be the space of $E$-valued, $\G$-predictable processes.

\section{Some motivation: one market-maker facing several market-takers}\label{sec_N_players}

In this section, we provide motivations for the use of mean-field games to model the interaction between a market-maker and a large number of informed market-takers. We first begin by describing the problem of $N$ market-takers facing one market-maker. Since the purpose of this section is purely motivational, we adopt a heuristic approach in the presentation, and avoid properly defining the weak formulation of the problem, which we reserve to the mean-field version in \Cref{weakformsec}. 

\subsection{Framework}

We consider a trading horizon $T>0$ and a filtered probability space $\big(\Omega,\mathcal F, \mathbb F := (\mathcal{F}_t)_{t\in [0,T]},\mathbb{P}\big)$ under which all the stochastic processes are defined. We set some positive integer $N\in \mathbb{N}^\star$, corresponding to the number of informed market-takers acting on a market consisting of a single asset whose price process is denoted by $S.$ This quantity, whose definition will be detailed later, can be thought as the mid-price of the asset which can be impacted by the trading activity of the market-takers. 

\medskip
We consider a market whose functioning is as follows: for any $i\in\{1,\dots,N\}$, at time $t \in [0,T]$, the $i$-th trader  has a fair buy or sell price in mind, around the mid-price $S_t$, meaning that he is ready to buy or sell the asset at time $t$ at prices respectively given by
\[
P^{b}(S_t,\delta^{i,b}_t):= S_t - \delta^{i,b}_t,\; \text{\rm and}\; 
P^{a}(S_t,\delta_t^{i,a}):= S_t + \delta^{i,a}_t,
\]
where for $i\in \{1,\dots,N\}$, $\big(\delta^{i,b},\delta^{i,a}\big)$ are what we call the control processes of the $i$-th trader. The $i$-th trader can trade with any participant of the market. His number of filled bid and ask orders are modelled by counting processes $N^{i,b}$ and $N^{i,a}$, with intensities respectively given by $\lambda^{i,b, \delta^i} := \Lambda^m ( \delta^{i,b} )$ and $\lambda^{i,a, \delta^i} := \Lambda^m (  \delta^{i,a})$, where
\begin{align}\label{eq:intensities}
&\Lambda^m( x):= A^m \exp \bigg( -\frac{k^m}{\sigma} x  \bigg), \; x\in\R.
\end{align}
\medskip
The form of the intensities deserves several comments, which hold for any $i\in\{1,\dots,N\}$
\begin{itemize}
    \item[$(i)$] the coefficient $A^m>0$ is the average number of transactions by unit of time when the $i$-th market-takers' fair price is identically equal to the mid-price;
    \item[$(ii)$] the coefficient $k^m>0$ represents the sensitivity of the intensity to the mid-to-bid and ask-to-mid quotes proposed by the $i$-th market-taker. More precisely, the market-taker trades frequently at the bid (resp. at the ask) when his mid-to-bid (resp. ask-to-mid) price is relatively small;
    \item[$(iii)$] the intensities are decreasing functions of the volatility of the asset $\sigma>0$, defined below, meaning that a high volatility induces a lower number of trades and conversely, see for example \citeauthor*{dayri2015large} \cite{dayri2015large}, \citeauthor*{madhavan1997security} \cite{madhavan1997security}, or \citeauthor*{wyart2008relation} \cite{wyart2008relation}.  
\end{itemize}

At any time $t\in[0,T]$, the market-maker, who does not know the fair prices of the market-takers, proposes quotes around the mid-price $S_t$ at which she  is ready to buy or sell the asset to the $i$-th trader. For $t\in[0,T]$, these prices are
\begin{align*}
    P^{b}(S_t,\delta_t^{0,b}) := S_t -\delta_t^{0,b}, \; \text{\rm and}\; P^{a}(S_t,\delta_t^{0,a}) := S_t + \delta_t^{0,a},\; t\in[0,T],
\end{align*}
where $(\delta^{0,b},\delta^{0,a})$ are what we call the control processes of the market-maker. The set of admissible controls for the $N+1$ players is defined as
\begin{align*}
    \mathcal{A}_N = \Big\{ (\delta^0, \delta^{-0}) = \big((\delta^{0,b},\delta^{0,a}),(\delta^{i,b},\delta^{i,a})_{i\in \{1,\dots,N\}}\big)\in\L^0\big( \Mc_{2,N+1}(\R),\F\big):  (\delta^0, \delta^{-0})\; \text{\rm bounded by}\; \delta_\infty\Big\},
\end{align*}
for a given $\delta_{\infty}>0$. The optimisation problem of the $i$-th market-taker is a function of the spread vector $\delta^{-i}$ quoted by the $N-1$ other market-takers and the spread $\delta^0$ of the market-maker. Hence, each market-taker will maximise his PnL given the actions of the other market-takers to obtain his so-called best-reaction function. We define the set of admissible controls of the $i$-th player given controls $\delta^0$ and $\delta^{-i}$ chosen by the other market participants by
\begin{align*}
\mathcal{A}_N^i(\delta^0, \delta^{-i}) := \Big\{\delta\in\L^0\big( \mathcal{M}_{2,1}(\mathbb{R}),\F\big):( \delta^0, \delta \otimes_i \delta^{-i}) \in \mathcal{A}_N\Big\},
\end{align*}
and the set of admissible controls of the market-maker given a control $\delta^{-0}$ played by other market participants
\begin{align*}
\mathcal{A}_N^
0(\delta^{-0}) = \Big\{\delta\in\L^0\big(\mathcal{M}_{2,1}(\mathbb{R}),\F\big): ( \delta,  \delta^{-0}) \in \mathcal{A}_N\Big\}. 
\end{align*}

The number of transactions at the bid and at the ask for the market-maker is modelled by two processes $N^{0,b}$ and $N^{0,a}$, with intensity processes $\lambda^{0,b, \delta}$ and $\lambda^{0,a, \delta}$, respectively given by
\[
\lambda^{0,b, \delta}_t := \Lambda\bigg( \frac 1N \sum_{i=1}^N \delta_{t}^{i,a}, \delta_t^{0,b} \bigg),\; \text{and} \; \lambda^{0,a, \delta}_t :=  \Lambda \bigg( \frac 1N \sum_{i=1}^N \delta_{t}^{i,b}, \delta_t^{0,a} \bigg),\; \forall t \in [0,T],
\]
where
\begin{align}\label{eq:intensitymaj}
&\Lambda( x,y):= A \exp \bigg( -\frac{k}{\sigma} ( x + y )  \bigg), \; (x,y)\in\R^2,
\end{align}

for positive constants $A$ and $k$. Hence, these intensities depend on the difference between the average fair price of the market-takers on the one hand, and the price proposed by the market-maker on the other hand: for instance, if the average fair price at which market-makers are ready to sell the asset is very large compared to the price at which the market-maker is ready to buy, the market-maker will not trade often. In a framework where all the processes $(\delta^{i,b},\delta^{i,a})$ are constant equal to zero, that is for all $t\in [0,T], \delta_t^{i,a}=\delta_t^{i,b}=0$, the market-takers are considered as `passive', and we recover the classical optimal market-making framework of \citeauthor*{avellaneda2008high} \cite{avellaneda2008high}, \citeauthor*{cartea2015algorithmic} \cite{cartea2015algorithmic}, and \citeauthor*{gueant2013dealing} \cite{gueant2013dealing}.

\medskip
The dynamics of the price process $S$ under which the market-maker and the $N$ market-takers compute the marked-to-market value of their inventory is defined as 
\begin{align}\label{eq_mid_price_impacted}
    \d S_t =  \sigma \d W_t + \frac{\kappa}{N} \sum_{i=1}^N\big(\d N^{i,b}_t - \d N^{i,a}_t \big),
\end{align}
where $\kappa$ and $\sigma$ are positive constants, $W$ is a one-dimensional Brownian motion, independent of $N^{0,b}$ and $N^{0,a}$. This equation deserves several comments
\begin{itemize}
    \item[$(i)$] $\sigma W$ represents the diffusion part of the asset's price, independent from the trading activity;
    \item[$(ii)$] the asset price is impacted by the average behaviour of the market-takers, through the so-called permanent market impact effect: when, on average, market-takers buy (resp. sell) the asset at a high rate, its price increases (resp. decreases).
\end{itemize}
Finally, we assume that each market-taker reacts to a specific signal on the price, modelled for the $i$-th market-taker, for any $i\in\{1,\dots,N\}$, by a process $b^i$, verifying 
\[
b^i_t:=N_t^{i,+} - N_t^{i,-},\; t\in[0,T],
\] where $N^{i,+}$ and $N^{i,-}$ are Poisson processes of intensities $\bar{\lambda}^+$ and $\bar{\lambda}^-$, independent of $(W,N^{0,b},N^{0,a}, N^{i,b},N^{i,a})$.\footnote{See \cite{huang2019mean} for a similar set up. In the literature, the drift $b$ is often taken with a dynamics of the type \[
\d b_t = -\theta b_t \d t + \eta \d B_t + \d N^+_t - \d N^-_t,
\] where $(\theta,\eta) \in [0,+\infty)^2$ and $B$ is some Brownian motion independent of everything else. Here we take $\eta = \theta = 0$ for the sake of simplicity: this allows to reduce our problem to a system of ODEs.} This signal process corresponds to the belief of the $i$-th market-taker concerning the drift of the asset, which can be modified by the information he acquires over time.

\subsection{The problem of the market-maker}

The cash process $X^0$ of the market-maker is naturally defined as follows
\begin{align*}
   X^0_t := X^0_0 +\int_0^tP^{a}(S_s,\delta_s^{0,a}) \d N^{0,a}_s -  \int_0^tP^{b}(S_s,\delta_s^{0,b}) \d N^{0,b}_s ,\; t\in[0,T],
\end{align*}
with $X^0_0\in  \R$ given. Her inventory process $q^0$ is given by
\begin{align*}
    \d q^0_t :=  \d N^{0,b}_t -  \d N^{0,a}_t,
\end{align*}
with $q^0_0\in \Z$ given. For any $(\delta^0,\delta^{-0})\in\mathcal{A}_N$, we define a probability measure $\mathbb{P}^{\delta^0, \delta^{-0}}$ under which 
\begin{itemize}
    \item for all $i\in \{1, \ldots, N\}$, $N^{i,b}$ and $N^{i,a}$ are point processes of intensity $\Lambda^m (  \delta^{i,a} ) $ and $\Lambda^m (  \delta^{i,a} )$, respectively;
    \item $N^{0,b}$ and $N^{0,a}$ are point processes of intensity $\Lambda\bigg( \frac 1N \sum_{i=1}^N \delta^{i,a}, \delta^{0,b} \bigg)$ and $\Lambda \bigg( \frac 1N \sum_{i=1}^N \delta^{i,b}, \delta^{0,a} \bigg)$, respectively;
    \item $W$ is a one-dimensional Brownian motion, independent from $(N^{i,b},N^{i,a})_{i \in \{1,\ldots,N\}}$;
    \item the price process $S$ has dynamics \eqref{eq_mid_price_impacted};
    \item all the processes $N^{i,+}$ and $N^{i,-}$ have the same respective intensities $\bar \lambda^+$ and $ \bar \lambda^-$.
\end{itemize}

We denote by $\mathbb{E}^{\delta^0, \delta^{-0}}$ the expectation under the probability measure $\mathbb{P}^{\delta^0, \delta^{-0}}$. The market-maker wishes to maximise the sum of her cash process and the marked-to-market value of her inventory $q^0 S$, while keeping the inventory process close to zero during the trading period. Therefore, she wants to solve
\begin{align}\label{Optimization problem major player}
   V^0(\delta^{-0}):=  \sup_{\delta^0 \in \mathcal{A}_N^0(\delta^{-0})}\mathbb{E}^{\delta^0,\delta^{-0}}\bigg[X^0_T + q^0_T S_T - \phi\sigma^2\int_0^T ( q^0_t)^2 \mathrm{d}t \bigg],
\end{align}
where $\phi>0$ is the risk-aversion parameter of the market-maker. 

The quadratic penalty aims at penalising the quadratic variation of the term
\begin{align*}
X^0_T + q^0_T S_T = X^0_0 + q^0_0S_0 &+\int_0^T \delta^{0,a}_t \d N^{0,a}_t + \int_0^T \delta^{0,b}_t \d N^{0,b}_t + \int_0^T q^0_t \frac{\kappa}{N} \sum_{i=1}^N\big(\d N^{i,b}_t - \d N^{i,a}_t \big)  + \int_0^T \sigma q^0_t \d W_t.
\end{align*}
Using this expression, \Cref{Optimization problem major player} actually boils down to 
\begin{align}\label{Simplified_opt_major}
    V^0(\delta^{-0})&=\sup_{\delta^0 \in \mathcal{A}_N^0(\delta^{-0})}\mathbb{E}^{\delta^0,\delta^{-0}}\Bigg[\int_0^T \bigg(  \delta^{0,a}_t \Lambda\bigg( \frac 1N \sum_{i=1}^N \delta_{t}^{i,b}, \delta_t^{0,a} \bigg) + \delta^{0,b}_t \Lambda \bigg( \frac 1N \sum_{i=1}^N \delta_{t}^{i,a}, \delta_t^{0,b} \bigg)\\
    &\qquad\qquad\qquad \qquad\qquad+ q^0_t \frac{\kappa}{N} \sum_{i=1}^N\big(\Lambda ^m(  \delta_{t}^{i,b}) - \Lambda^m (\delta_{t}^{i,a}) \big)  - \phi\sigma^2 ( q^0_t)^2 \bigg) \mathrm{d}t \Bigg].\nonumber
\end{align}
Notice in particular that when $\delta^{-0}$ is given, the only state process of this problem is $q^0$. This trick, already used in \citeauthor*{bergault2021size} \cite{bergault2021size}, makes the problem mathematically and numerically more tractable, because the associated PDE boils down to a simple ODE.

\subsection{The problem of the market-takers} 

Each market-taker manages his inventory process $q^{i}$, $i\in\{1,\dots,N\}$, which is defined as the difference between the number of filled bid and ask requests
\begin{align}\label{eq_inventory_trader_i}
    q_t^{i}
    := N_t^{i,b} - N_t^{i,a},\; \forall t \in [0,T].
\end{align}
We define the dynamics of the cash process of the $i$-th trader $X^i$ by
\begin{align*}
     X^i_t :=X^i_0+\int_0^t  P^{a}(S_s,\delta_s^{i,a}) \d N^{i,a}_s  -\int_0^tP^{b}(S_s,\delta_s^{i,b}) \d N^{i,b}_s,\; t\in[0,T].
\end{align*}
The $i$-th market-taker observes a specific signal $b^i$, meaning that between time $t$ and time $t+\d t$, he expects to make an instantaneous additional gain (or loss) of $b^i_t q^i_t \d t.$ Given the quotes $\delta^0$ of the market-maker and $\delta^{-i}$ of the $N-1$ other market-takers, the $i$-th market-taker wants to solve
\begin{align}\label{Optimization problem minor player}
       V^i(\delta^0,\delta^{-i}):= \sup_{\delta^i \in \mathcal{A}_N^i(\delta^0,\delta^{-i})}\mathbb{E}^{\delta^0,\delta^i\otimes_i \delta^{-i}}\bigg[X_T^i + q_T^i S_T + \int_0^T b^i_tq^i_t \d t - \gamma \sigma^2 \int_0^T (q^i_t)^2 \d t \bigg],
\end{align}
where $\gamma>0$ is the risk-aversion parameter of the traders, assumed here to be homogeneous. The market-taker wants to maximise the sum of his expected cash process and the marked-to-market value of his inventory $q^i S$. The last term is a running penalty for the risk carried by the market-taker.

\medskip
As before, we can see that the problem boils down to 
\begin{align}\label{Simplified_opt_minor}
  \nonumber  V^i(\delta^0,\delta^{-i})&=\sup_{\delta^i \in \mathcal{A}_N^i(\delta^0,\delta^{-i})}\mathbb{E}^{\delta^0,\delta^i\otimes_i \delta^{-i}}\Bigg[ \int_0^T \bigg(\delta^{i,a}_t \Lambda^m (\delta^{i,a}_t ) + \delta^{i,b}_t \Lambda^m (\delta^{i,b}_t ) +  b^i_tq^i_t\\
&\qquad\qquad\qquad\qquad\qquad\qquad\quad+ q^i_t \frac{\kappa}{N} \sum_{j=1}^N\big(\Lambda ^m(  \delta_{t}^{j,b}) - \Lambda^m (\delta_{t}^{j,a}) \big) - \gamma \sigma^2 \int_0^T (q^i_t)^2 \bigg)  \d t \Bigg].
\end{align}
With $\delta^0$ and $\delta^{-i}$ given, the only state processes of this problem are $q^i$ and $b^i$.

\subsection{Limitations of the finite game model and mean-field limit}\label{section_comments_limitations}

In this section, we described an agent-based model where $N$ market-takers trade against one market-maker. Solving this multi-agent problem boils down to the resolution of a system of Hamilton--Jacobi--Bellman equations, where the state variables are the inventory processes of the market-maker and the $N$ market-takers. This system of $N+1$ HJB equations of dimension $N+2$ (that is, the time, the inventories of the $N$ market-takers and the inventory of the market-maker) is obviously intractable in practice for a large number of market-takers.

\medskip
In order to propose a tractable framework for the optimal market-making problem with strategic market-takers, we propose a mean-field game approach. The market-maker does not face $N$ strategic market-takers but infinitely many of them in a mean-field interaction, which can be thought as the averaged behaviour of the market-takers. In the next section, we present rigorously the mean-field limit of the $N$-players model, and the corresponding optimisation problems of the market-maker and the mean-field agent. 

\medskip
We see in \eqref{Simplified_opt_major} and \eqref{Simplified_opt_minor} that the empirical distribution of the controls of the market-takers appears in both problems. Though we do not pursue rigorous proofs in that direction, a typical reasoning using propagation of chaos (see for instance \citeauthor*{bayraktar2021mean} \cite{bayraktar2021mean}) leads to conjecture that as $N$ goes to $+\infty$, the controls $(\delta^{i,b}, \delta^{i,a})_{i \in \{1, \ldots, N \}}$ chosen by the market-takers will only intervene through their distribution, which we denote by $\nu$, \emph{i.e.} $\nu = (\nu_t)_{t \in [0,T]}$ denotes the probability flow representing the distribution of those controls on $\mathbb R^2$. For $t\in [0,T]$, we denote by $\nu^b_t$ and $\nu^a_t$ the first and second marginal laws of $\nu_t$, respectively. 
\medskip

Propagation of chaos also tells us that at the limit, we should expect that the point processes associated to each market-taker will become independent, so that it is reasonable to consider that the intensities of  $N^{0,b} $ and $N^{0,a} $ will respectively converge towards
\[
A \exp \bigg( - \frac{k}{\sigma} \bigg( \delta^{0,b}_t +  \int_{\mathbb R} x  \nu^a_t(\mathrm{d}x)\bigg) \bigg),\; \text{and} \; A  \exp \bigg( - \frac{k}{\sigma } \bigg( \delta^{0,a}_t +  \int_{\mathbb R} x \nu^b_t(\mathrm{d}x)  \bigg) \bigg),\; t\in[0,T] .\]
The intensities of the processes $N^{i,b}$ and $N^{i,a}$ of a given market-taker do not change, as they do not depend on the controls of the other market-takers
\[\lambda^{i,b,\delta^i}_t = A^m  \exp \bigg( -\frac{k^m}{\sigma} \delta^{i,b}_{t}  \bigg),\; \lambda^{i,a,\delta^i}_t =A^m  \exp \bigg( -\frac{k^m}{\sigma} \delta^{i,a}_{t}  \bigg),\; t\in[0,T].\]

In the next section, we rigorously introduce the mean-field version of the problem.

\section{The mean-field problem}
\label{mfg_proba_framework}
\subsection{Probabilistic framework}

\label{weakformsec}

Let $T>0$ be a final horizon time, $\Omega_d$ the set of piecewise constant càdlàg functions from $[0,T]$ into $\mathbb N$, and $\Omega := \Omega_d ^6$. The observable state is the canonical process $\chi := \big( N^{b},   N^{a}, N^{m,b},   N^{m,a},  N^+, N^- \big)$ of the space $\Omega$, with
\[
 N^{b}_t(\omega) =  n^{b}(t), \;  N^{a}_t(\omega) = n^{a}(t),
\]
\[
N^{m,b}_t(\omega) =  n^{m,b}(t), \; N^{m,a}_t(\omega) =  n^{m,a}(t), \;N^+_t(\omega) = n^+(t), \; N^-_t(\omega) = n^-(t), 
\]
for all $t \in [0,T]$, and $\omega = (n^{b}, n^{a}, n^{m,b}, n^{m,a}, n^+, n^-) \in \Omega.$ We introduce positive constants $(A, A^m, \bar \lambda, \tilde q, \tilde q^m, \tilde b) $, as well as the unique probability\footnote{The existence and uniqueness of $\mathbb P$ is proved for example in \citeauthor*{el2021optimal} \cite[Lemma A.1]{el2021optimal}} $\mathbb P$ on $\Omega$ such that, under $\mathbb P$, $W$ is a standard Brownian motion, the two point processes $N^{b}$ and $N^{a}$ have respective intensities\footnote{We now impose risk limits for the market-maker and the representative market-taker, in terms of boundaries for the processes $N^b - N^a$ and $N^{m,b}-N^{m,a}$. Similarly, we bound the process $N^+ - N^-$. This is mainly for technical reasons, in order to simplify the problem. The boundaries $\tilde q^m$ and $\tilde b$ are assumed to be the same for all the market-takers.}
\[
\lambda^{b}_t := A \mathds 1_{\{ N^{b}_{t-} -  N^{a}_{t-} < \tilde q\}},\; \lambda^{a}_t := A \mathds 1_{\{  N^{b}_{t-} -  N^{a}_{t-} >- \tilde q\}},\; t\in[0,T],
\]
the two point processes $N^{m,b}$ and $N^{m,a}$ have respective intensities
\[
\lambda^{m,b}_t := A^m \mathds 1_{\{ N^{m,b}_{t-} -  N^{m,a}_{t-} < \tilde q^m\}},\; \lambda^{m,a}_t := A^m \mathds 1_{\{  N^{m,b}_{t-} -  N^{m,a}_{t-} >- \tilde q^m\}},\; t\in[0,T],
\]
 the two point processes $N^+$ and $N^-$ have respective intensities
\[
\lambda^+_t = \bar \lambda^+ \mathds 1_{\{  N^+_{t-} -  N^-_{t-} < \tilde b\}},\; \lambda^-_t = \bar \lambda^- \mathds 1_{\{  N^+_{t-} -  N^-_{t-} >- \tilde b\}}, \; t\in[0,T],
\]
and the processes $(N^{b}, N^{a})$, $(N^{m,b}, N^{m,a})$, $(N^+,N^-)$ are $\P$-independent. We denote by $q := N^{b} - N^{a}$ the inventory process of the market-maker, and by $q^m: = N^{m,b} - N^{m,a}$ the inventory process of the representative market-taker. We also denote by $b^m: = N^+ - N^-$ the signal observed by the representative market-taker for the asset. Finally, we denote the canonical $\mathbb P$-completed filtration generated by $\chi$ by $\mathbb F := (\mathcal F_t)_{t\in [0,T]}$ .

\subsection{Admissible controls and changes of measure}

Let us define, for a given $\delta_\infty >0$, the set of admissible controls
\[
\mathcal A_\infty = \Big\{ (\delta^{b}, \delta^{a})\in\L^0(  \mathbb R^2,\F): (\delta^{b}, \delta^{a}) \text{ \rm bounded by} \; \delta_\infty \Big\}.
\]

As explained before, the market-maker chooses the process $(\delta^{b},\delta^{a}) \in \mathcal A_\infty$ to fix the bid and ask prices respectively given by
\[
P^{b}(S_t,\delta^{b}) = S_t - \delta^{b}_t, \; \text{and} \; P^{a}(S_t,\delta^{a}) = S_t + \delta^{a}_t, \; t\in[0,T].
\]

Similarly, the representative market-taker chooses a process denoted by $(\delta^{m,b},\delta^{m,a}) \in \mathcal A_\infty$ to fix what he considers to be the `fair' bid and ask prices
\[
 P^{b}(S_t,\delta_t^{m,b}) = S_t -\delta^{m,b}_t, \; \text{and} \; P^{a}(S_t, \delta_t^{m,a}) = S_t + \delta^{m,a}_t, \; t\in[0,T].
 \]

Let us denote by $\mathcal P (\mathbb R^4)$ the space of probability measures on $\mathbb R^4$. We denote by $\Dc^p(\mathbb R^4)$ the space of deterministic measurable functions from $[0,T]$ to $\mathcal P(\mathbb R^4),$ and introduce the probability flow $\mathcal{V} \in \Dc^p(\mathbb R^4)$ representing the joint distribution of the controls of the mean-field of market-takers, their inventories, and their specific signals. For $t \in [0,T]$, we denote by $\nu^b_t$ and $\nu^a_t$ the first and second marginal laws of $\mathcal{V}_t,$ respectively, representing the distributions of the controls at the bid and at the ask. We denote by $\mu_t$ the third marginal, representing the distribution of the inventories of the mean-field of market-takers, and we denote by $\beta_t$ the fourth marginal, representing the distribution of the signals they observe.\medskip

Let us define the following exponential operator
\begin{align*}
   \bar{\mathcal{E}}_{\beta}(N)_T := \exp\bigg(\int_0^T \log(1+\beta_s)\d N_s -\int_0^T \beta_s \lambda_s \d s\bigg),
\end{align*}
for a point process $N$ with intensity process $\lambda$ and a process $\beta >-1$, $\P$--a.s. We finally introduce the probability measure $\mathbb P^{\delta, \delta^m,\Vc}$ given by
\begin{align}
\label{probaMM}
 \frac{\d\mathbb P^{\delta, \delta^m,\Vc}}{\d \mathbb P}  :=\bar{\mathcal{E}}_{\beta^{ \delta, \nu, b}}(N^{b})_T
 \bar{\mathcal{E}}_{\beta^{ \delta, \nu, a}}(N^{a})_T \bar{\mathcal{E}}_{\beta^{  \delta^m, b}}(N^{m,b})_T
 \bar{\mathcal{E}}_{\beta^{\delta^m, a}}(N^{m,a})_T, 
\end{align}
where for $j\in\{a,b\}$
\begin{align*}
\beta^{ \delta, \nu, b}_t := \frac 1{A}  {\Lambda}\bigg(  \int_{\mathbb R}x \nu^a_t(\mathrm{d}x) , \delta^{b}_t\bigg) - 1, \; \beta^{ \delta, \nu, a}_t := \frac 1{A}  {\Lambda}\bigg(  \int_{\mathbb R} x\nu^b_t(\mathrm{d}x) , \delta^{a}_t\bigg)   - 1, \; \beta^{  \delta^m, j}_t := \frac 1{A^m}  {\Lambda}^m(  \delta^{m,j}_t) - 1,\; t\in[0,T].
\end{align*}
We denote by $\mathbb E^{\delta, \delta^m, \Vc}$ the expectation under $\mathbb P^{\delta, \delta^m, \Vc}$. To sum up, under $\mathbb P^{ \delta, \delta^m, \Vc}$, $N^{b}$ and $N^{a}$ have respective intensities given by
\[
\lambda^{b,\delta,\nu}_t = \mathds 1_{ \{ q_{t-} <\tilde q\}}   {\Lambda}\bigg (\int_{\mathbb R} x \nu^a_t(\mathrm{d}x), \delta^{b}_t \bigg)  ,\; \text{and} \; \lambda^{a,\delta,\nu}_t = \mathds 1_{ \{ q_{t-} >-\tilde q \}} {\Lambda}\bigg( \int_{\mathbb R} x\nu^b_t(\mathrm{d}x) , \delta^{a}_t\bigg) ,\; t\in[0,T],
\]
 $N^{m,b}$ and $N^{m,a}$ have  intensities $\lambda^{m,b}$ and $\lambda^{m,a}$ respectively given by
\[\lambda^{m,b, \delta^m}_t := \mathds 1_{ \{ q^m_{t-} <\tilde q^m\}}  {\Lambda}^m ( \delta^{m,b}_{t} ), \; \text{and}\; \lambda^{m,a, \delta^m}_t := \mathds 1_{ \{ q^m_{t-} >-\tilde q^m\}}{\Lambda}^m ( \delta^{m,a}_{t}),\; t\in[0,T],
\]
while $N^+$ and $N^-$ still have respective intensities $\bar \lambda^+ \mathds 1_{\{  N^+_{t-} -  N^-_{t-} < \tilde b\}}$, and $\bar \lambda^- \mathds 1_{\{  N^+_{t-} -  N^-_{t-} >- \tilde b\}}, \;t\in[0,T].$

\begin{remark}
Note that, to ease the notations, we only consider symmetric intensities $($same $\Lambda$ and $\Lambda^m$ at the bid and at the ask$)$ and symmetric risk limits. The generalisation to the asymmetric case, with different flows at the bid and at the ask, is of course straightforward. 
\end{remark}

\subsection{Optimisation problems and definition of equilibria}

Let us define the function $\Lc: \llbracket -\tilde q^m, \tilde q^m \rrbracket \times \mathbb R^2 \longrightarrow \mathbb R$ by $\Lc(y,x^b,x^a):=\mathds1_{ \{y>-\tilde q^m\}}\Lambda^m(x^b) - \mathds 1_{ \{y <\tilde q^m\}}\Lambda^m(x^a).$ The inventory process $q$ of the market-maker has dynamics $\d q_t = \d N^{b}_t - \d N^{a}_t,$ with $q_0 \in \llbracket - \tilde q, \tilde q \rrbracket$. Taking the mean-field version of  \eqref{Simplified_opt_major}, we consider the following problem for the market-maker
\begin{align}\label{major_value_function}
\begin{split}
    V( \Vc) &:= \sup_{\delta\in\mathcal A_\infty}\mathbb E^{\delta, \delta^m, \Vc} \bigg[ \int_0^T \bigg(  \delta^{b}_s  \mathds{1}_{\{ q_{s-} < \tilde q\}} {\Lambda}\bigg(\int_{\mathbb R} x \nu^a_s(\mathrm{d}x), \delta^{b}_s\bigg)   + \delta^{b}_s \mathds{1}_{\{ q_{s-} > - \tilde q\}} {\Lambda} \bigg( \int_{\mathbb R} x\nu^b_s(\mathrm{d}x), \delta^{a}_s\bigg)   \\ 
    &\qquad\qquad\qquad\qquad\quad - \phi \sigma^2 (q_s)^2  +  q_s \kappa \int_{\mathbb R^4}   \Lc (y,  x^b,  x^a  ) \Vc_s(\mathrm{d}x^b, \mathrm{d}x^a, \d y, \d z)  \bigg)\d s  \bigg],
\end{split}
\end{align}
for a given probability flow $\Vc \in \Dc^p(\mathbb R^4)$. We now see that $q$ is stil the only state variables of this problem. In particular, we observe that this problem is actually independent of $\delta^m$, as the process $q$ and the market impact term only depend on the average behaviour of the market-takers represented by the probability flow $\mathcal V$, but not on the behaviour of the representative market-taker alone. This is why we did not spell out any explicit dependent of $V$ on $\delta^m$.

\medskip
Next, the inventory process $q^m$ of the market-taker has dynamics $\d q^m_t = \d N^{m,b}_t - \d N^{m,a}_t,$ with $q^m_0 \in \llbracket - \tilde q^m, \tilde q^m \rrbracket$. Taking the mean-field version of problem \eqref{Simplified_opt_minor}, we consider the following problem for the representative market-taker
\begin{align}
\begin{split}
   V^m(\Vc)  &:= \sup_{\delta^m \in \mathcal{A}_\infty}\mathbb E^{\delta, \delta^m, \Vc} \bigg[  \int_0^T \bigg(  \delta^{m,b}_s\mathds{1}_{\{ q^m_{s-} < \tilde q^m\}}  \Lambda ^m( \delta^{m,b}_{t} ) + \delta^{m,a}_s  \mathds{1}_{\{ q^m_{s-} >- \tilde q^m\}}  \Lambda^m ( \delta^{m,a}_{t} )) \\
    &\qquad\qquad\qquad\qquad\qquad  + b^m_s q^m_s  - \gamma \sigma^2 (q^m_s)^2 +  q^m_s \kappa  \int_{\mathbb R^4} \Lc (y, x^b, x^a )  \Vc_s( \mathrm{d}x^b, \mathrm{d}x^a, \d y, \d z)\bigg)\d s   \bigg],    
\end{split}
\end{align}
again for a given probability flow $\Vc\in \Dc^p(\mathbb R^4)$. We now see that the only two state variables to this problem are $b^m$ and $q^m$. In particular, as before, we observe that, due to our choice of intensity functions, the problem of the representative market-taker does not depend on $\delta$. Finally, we can now define a solution to the mean-field game.
\begin{definition}\label{geneq}
A solution of the above game is given by a probability flow $\Vc^{\star}\in \Dc^p(\mathbb R^4)$, a control $ \delta^{\star}\in \mathcal A_\infty$, and a control $\delta^{m,\star}\in \mathcal A_\infty$ such that
\begin{itemize}
    \item[$(i)$] the control $\delta^{\star}$ reaches the supremum in the definition of $ V(\Vc^{\star});$
    \item[$(ii)$] the control $\delta^{m,\star}$ reaches the supremum in the definition of ${V}^m( \Vc^{\star});$
    \item[$(iii)$] $\Vc^{\star}_t$ is the joint distribution of $\delta^{m,\star}_{t},$ $q^m_{t},$ and $b^m_{t}$ under ${\mathbb P}^{ \delta^\star, \delta^{m,\star},\Vc^\star }$ for Lebesgue--almost every $t\in[0,T]$.
\end{itemize}
\end{definition}

\section{A Markovian version of the MFG problem}
\label{MarkovCasesec}

To ease the above problem, among all equilibria, we will only study here the Markovian ones. This approach is inspired by \citeauthor*{carmona2017alternative} \cite{carmona2017alternative}. In our case, this allows to write a numerically tractable system of PDEs. 

\subsection{Admissible controls and change of measure}

Let us denote by $\mathcal P (\mathbb R^2)$ the space of probability measures on $\mathbb R^2$. Let us also denote by $\Dc$ the space of deterministic measurable functions from $[0,T] \times  [-\tilde q, \tilde q] \times \mathcal P (\mathbb R^2)$ to $[-\delta_{\infty}, \delta_{\infty}]$, and by $\Dc^m$ the space of deterministic measurable functions from $[0,T]\times [-\tilde b, \tilde b]\times  [-\tilde q^m, \tilde q^m] \times \mathcal P (\mathbb R^2)$  to $[-\delta_{\infty}, \delta_{\infty}]$. Finally, we denote by $\Dc^p(\mathbb R^2)$ the space of deterministic measurable functions from $[0,T]$ to $\mathcal P(\mathbb R^2).$ We now define the set of admissible controls for the market-maker as
\begin{align*}
 \bar{\mathcal A}_\infty( \Pi) &= \Big\{ (\delta^{b}, \delta^{a})\in\Ac_\infty: \exists  (\bar \delta^b, \bar \delta^a) \in \Dc^2,\; \delta^i_t = \bar \delta^b (t, q_{t-}, \Pi_{t}), \;   \mathbb P \otimes \mathrm{d}t\text{\rm--a.e.},\; i\in\{a,b\} \Big\},
\end{align*}
where the probability flow $\Pi = (\Pi_t)_{t \in [0,T]} \in \Dc^p(\mathbb R^2)$ represents the joint distribution of the inventories of the mean-field of market-takers and the signals they observe. Similarly, the admissible controls for the representative market-taker now lie in the set
\begin{align*}
\bar{\mathcal A}_\infty^m(\Pi) &= \Big\{ (\delta^{m,b}, \delta^{m,a})\in\Ac_\infty: \exists  (\bar \delta^{m,i}, \bar \delta^{m,a}) \in (\Dc^m)^2,\; \delta^{m,i}_t = \bar \delta^{m,i} (t,b^m_{t-},q^m_{t-}, \Pi_{t}),\;   \mathbb P \otimes \mathrm{d}t\text{\rm--a.e.}, i\in\{a,b\} \Big\}.
\end{align*}

For a probability flow $\Pi$ on $\mathbb R^2$, considering that all the market-takers choose their controls according to a Markovian function $\bar \delta^{\text{\fontsize{4}{4}\selectfont \rm MF}} = \big(\bar \delta^{\text{\fontsize{4}{4}\selectfont \rm MF},b}, \bar \delta^{\text{\fontsize{4}{4}\selectfont \rm MF},a} \big) \in (\Dc^m)^2$, the probability flow $\Vc$ on $\mathbb R^4$ of the distribution of the controls, the inventories and the signals of the market-takers, is entirely determined by $\bar \delta^{{\text{\fontsize{4}{4}\selectfont \rm MF}}}$ and $\Pi$. Hence, we will denote it $\Vc \big(\bar \delta^{{\text{\fontsize{4}{4}\selectfont \rm MF}}}, \Pi \big)$. Under ${\mathbb P}^{\delta, \delta^m,\Vc (\bar \delta^{{\text{\fontsize{4}{4}\selectfont \rm MF}}}, \Pi )}$, $N^+$, $N^-$ have intensities $\bar \lambda^+ \mathds 1_{\{  N^+_{t-} -  N^-_{t-} < \tilde b\}}$, $\bar \lambda^- \mathds 1_{\{  N^+_{t-} -  N^-_{t-} >- \tilde b\}}$, those of $N^{b}$ and $N^a$ are
\[
\mathds 1_{ \{ q_{t-} <\tilde q\}}   {\Lambda}\bigg ( \int_{\mathbb R^2}\bar \delta^{{\text{\fontsize{4}{4}\selectfont \rm MF}},a} (t,z,y, \Pi_{t}) \Pi_t(\mathrm{d}y, \d z) , \delta^{b}_t \bigg)  ,\; \text{and} \; \mathds 1_{ \{ q_{t-} >-\tilde q \}} {\Lambda}\bigg( \int_{\mathbb R^2}\bar \delta^{{\text{\fontsize{4}{4}\selectfont \rm MF}},b} (t,z,y, \Pi_{t}) \Pi_t(\mathrm{d}y, \d z), \delta^{a}_t\bigg) ,\; t\in[0,T].
\]
and those of $N^{m,b}$ and $N^{m,a}$ are given by $\lambda^{m,b,\delta^m}_t= \mathds 1_{ \{ q^m_{t-} <\tilde q^m\}}  {\Lambda} ( \delta^{m,b}_{t} ) $
and
$\lambda^{m,a,\delta^m}_t = \mathds 1_{ \{ q^m_{t-} >-\tilde q^m\}}{\Lambda} ( \delta^{m,a}_{t})$, $t\in[0,T]$. 

\subsection{The Markovian optimisation problems}

With the same notations as before, the optimisation problem of the market-maker is now given by
\begin{align}\label{major_value_function_markov}
\begin{split}
    \tilde V(\bar \delta^{{\text{\fontsize{4}{4}\selectfont \rm MF}}}, \Pi) &= \sup_{\delta\in\tilde {\mathcal A}_\infty(\Pi)} {\mathbb E}^{\delta, \delta^m,\Vc (\bar \delta^{{\text{\fontsize{4}{4}\selectfont \rm MF}}}, \Pi )} \Bigg[ \int_0^T \bigg(  \delta^{b}_s  \mathds{1}_{\{ q_{s-} < \tilde q\}} {\Lambda} \bigg( \int_{\mathbb R^2}\bar \delta^{{\text{\fontsize{4}{4}\selectfont \rm MF}},a} (s,z,y,  \Pi_{s})\Pi_s(\mathrm{d}y, \d z) , \delta^{b}_s\bigg)  \\
    &\qquad\qquad\qquad\qquad\qquad\qquad+ \delta^{a}_s \mathds{1}_{\{ q_{s-} > - \tilde q\}} {\Lambda} \bigg( \int_{\mathbb R^2}\bar \delta^{{\text{\fontsize{4}{4}\selectfont \rm MF}},b} (s,z,y,  \Pi_{s})\Pi_s(\mathrm{d}y, \d z) , \delta^{a}_s\bigg) - \phi \sigma^2 (q_s)^2 \\
    &\qquad\qquad\qquad\qquad \qquad\qquad+  q_s \kappa \int_{\mathbb R^2}   \Lc \big(y,  \bar \delta^{{\text{\fontsize{4}{4}\selectfont \rm MF}},b} (s,z,y, \Pi_{s}),  \bar \delta^{{\text{\fontsize{4}{4}\selectfont \rm MF}},a} (s,z,y,  \Pi_{s}) \big) \Pi_s(\mathrm{d}y, \d z)  \bigg)\d s  \Bigg],
\end{split}
\end{align}
for given $ (\bar \delta^{{\text{\fontsize{4}{4}\selectfont \rm MF}}} ,\Pi)\in (\Dc^m)^2\times\Dc^p(\mathbb R^2)$. Similarly, the optimisation problem of the representative market-taker is now
\begin{align}
\begin{split}
   \tilde {V}^m(\bar \delta^{{\text{\fontsize{4}{4}\selectfont \rm MF}}}, \Pi) & = \sup_{\delta^m \in \tilde {\mathcal{A}}_\infty^m(\Pi)}{\mathbb E}^{\delta, \delta^m,\Vc (\bar \delta^{{\text{\fontsize{4}{4}\selectfont \rm MF}}}, \Pi )} \Bigg[  \int_0^T \bigg( \delta^{m,b}_s\mathds{1}_{\{ q^m_{s-} < \tilde q^m\}}  \Lambda^m ( \delta^{m,b}_{s} ) + \delta^{m,a}_s  \mathds{1}_{\{ q^m_{s-} >- \tilde q^m\}}  \Lambda ^m( \delta^{m,a}_{s} )+ b^m_s q^m_s\\
    &\qquad\qquad    - \gamma \sigma^2 (q^m_s)^2  +  q^m_s \kappa  \int_{\mathbb R^2}   \Lc \big(y,  \bar \delta^{{\text{\fontsize{4}{4}\selectfont \rm MF}},b} (s,z,y, \Pi_{s}),  \bar \delta^{{\text{\fontsize{4}{4}\selectfont \rm MF}},a} (s,z,y,  \Pi_{s}) \big) \Pi_s(\mathrm{d}y, \d z)  \bigg)\bigg)\d s   \Bigg],    
\end{split}
\end{align}

for a given $ \bar \delta^{{\text{\fontsize{4}{4}\selectfont \rm MF}}} \in (\Dc^m)^2$ and $\Pi\in \Dc^p(\mathbb R^2)$. We now define a solution to the mean-field game in the Markovian case.

\begin{definition}\label{markoveq}
A solution of the above game in the Markovian case is given by functions $\bar \delta^{\star} = (\bar \delta^{\star,b}, \bar \delta^{\star,a}) \in \Dc ^2$, $\bar \delta^{m,\star} = (\bar \delta^{m,\star,b}, \bar \delta^{m,\star,a}) \in (\Dc^m )^2$, $\bar \delta^{{\text{\fontsize{4}{4}\selectfont \rm MF}},\star} = (\bar \delta^{{\text{\fontsize{4}{4}\selectfont \rm MF}},\star,b}, \bar \delta^{{\text{\fontsize{4}{4}\selectfont \rm MF}},\star,a}) \in (\Dc^m )^2$, and a probability flow $\Pi^{\star} \in \Dc^p(\mathbb R^2)$, such that
\begin{itemize}
    \item [$(i)$] the control $\delta^{\star}_t := \bar \delta^{\star}(t,  q_{t-}, \Pi^\star_t), \; \forall t \in [0,T]$, reaches the supremum in the definition of $\tilde V(\bar \delta^{{\text{\fontsize{4}{4}\selectfont \rm MF}},\star}, \Pi^{\star});$
        \item [$(ii)$]  the control $\delta^{m,\star}_t := \bar \delta^{m,\star}(t, b^m_{t-},q^m_{t-},\Pi^\star_t), \; \forall t \in [0,T]$, reaches the supremum in the definition of $\tilde {V}^m( \bar \delta^{{\text{\fontsize{4}{4}\selectfont \rm MF}},\star}, \Pi^{\star});$
        \item [$(iii)$]  $\bar \delta^{m,\star} = \bar \delta^{{\text{\fontsize{4}{4}\selectfont \rm MF}},\star};$
        \item [$(iv)$]  $\Pi^{\star}_t$ is the joint distribution of $q^m_{t}$ and $b^m_{t}$ under ${\mathbb P}^{ \delta^\star, \delta^{m,\star},\Vc (\bar \delta^{{\text{\fontsize{4}{4}\selectfont \rm MF}}, \star}, \Pi^\star )}$ for Lebesgue--almost every $t\in[0,T]$.
\end{itemize}
\end{definition}

\begin{remark}
Note that there are four items in this definition of a Markovian equilibrium, unlike in {\rm\Cref{geneq}}. This is because the probability flow $ \mathcal V^\star$ appearing in {\rm\Cref{geneq}} is now characterised by the pair $\big(\bar \delta^{{\text{\fontsize{4}{4}\selectfont \rm MF}},\star},  \Pi^{\star}\big)$.
\end{remark}

\subsection{The master equation}

We define the function $H: [-\delta_\infty,\delta_\infty] \times \mathbb{R}\times\mathbb{R}\longrightarrow \mathbb{R}$ as $H(d,z,\tilde{z}) = d  + \tilde{z} - z,$
and the function $\mathcal{H}:[0,T] \times \llbracket - \tilde q, \tilde q \rrbracket \times [-\delta_\infty,\delta_\infty]^2 \times \mathbb{R}\times\mathbb{R}\times\mathbb{R} \times (\Dc^m)^2 \times \Pc(\mathbb R^2)\longrightarrow \mathbb{R}$
\begin{align}\label{HamMM}
 \nonumber   \mathcal{H}(t,q,d_1,d_2,z,\tilde{z}_1,\tilde{z}_2,\bar \delta^{{\text{\fontsize{4}{4}\selectfont \rm MF}}},\Pi_t)&:= \mathds{1}_{\{ q <\tilde q\}} {\Lambda} \bigg(\int_{\mathbb R^2}\bar \delta^{{\text{\fontsize{4}{4}\selectfont \rm MF}},a}(t,z,y, \Pi_t) \Pi_t(\mathrm{d}y, \d z), d_1\bigg)  H\big(d_1,z,\tilde{z}_1\big) \\
    &\quad+ \mathds{1}_{\{ q > - \tilde q\}} {\Lambda} \bigg(\int_{\mathbb R^2}\bar \delta^{{\text{\fontsize{4}{4}\selectfont \rm MF}},b}(t,z,y, \Pi_t) \Pi_t(\mathrm{d}y, \d z), d_2\bigg) H\big(d_2,z,\tilde{z}_2\big).
\end{align}
This function corresponds to the Hamiltonian of the market-maker's problem: it accounts for the changes in the value function at each trade, as well as the marked-to-market gain associated with each trade. The first and second terms of the sum correspond respectively to the bid and ask side of the market-maker. We next introduce the function $\Kc : \llbracket - \tilde b , \tilde b \rrbracket  \times \mathbb{R} \times \mathbb{R} \times \mathbb{R} \longrightarrow \mathbb{R}$ defined by
\begin{align*}
    \Kc(b^m,z,\tilde{z}_1,\tilde{z}_2) =  \mathds{1}_{\{b^m< \tilde b\}} \bar \lambda^+ (  \tilde{z}_1 -  z ) + \mathds{1}_{\{b^m>- \tilde b\}} \bar \lambda^- (  \tilde{z}_2 -  z  ),
\end{align*}
corresponding to the infinitesimal generator of $b^m$. We define the function $\mathcal{H}^m:[0,T] \times \llbracket - \tilde b , \tilde b \rrbracket \times \llbracket - \tilde q^m , \tilde q^m \rrbracket \times  [-\delta_\infty,\delta_\infty]^2 \times \mathbb{R}\times\mathbb{R}\times \mathbb{R}\longrightarrow \mathbb{R}$ as 
\begin{align}\label{HamMT}
    \mathcal{H}^m(q^m,d_1,d_2,z,\tilde{z}_1,\tilde{z}_2) & = \mathbf{1}_{\{q^m<\tilde{q}^m\}}\Lambda^m ( d_1) H (d_1, z, \tilde z_1)+ \mathbf{1}_{\{q^m>-\tilde{q}^m\}}\Lambda^m( d_2) H (d_2, z, \tilde z_2).
\end{align}
As for the market-maker, this function corresponds to the Hamiltonian of the representative market-taker's problem.\medskip

\begingroup
\allowdisplaybreaks
Given the control function $\bar \delta^{{\text{\fontsize{4}{4}\selectfont \rm MF}}} = (\bar \delta^{{\text{\fontsize{4}{4}\selectfont \rm MF}},b}, \bar \delta^{{\text{\fontsize{4}{4}\selectfont \rm MF}},a})\in (\Dc^m)^2$ of the market-takers, if we denote by $p : [0,T] \times \llbracket - \tilde q^m , \tilde q^m \rrbracket \times \llbracket - \tilde b , \tilde b \rrbracket \longrightarrow \mathbb R$ the function such that $p(t,\cdot,\cdot)$ represents the probability mass function of the distribution $\Pi_t$, then $p$ verifies, at least formally, the following Fokker--Planck equation
\begin{align}\label{FP}
\left\{\begin{aligned}
 & \partial_t p(t,q^{m},b^m) - \mathds{1}_{\{ q^{m}>- \tilde q^m\}} \Lambda^m  \big(\bar \delta^{{\text{\fontsize{4}{4}\selectfont \rm MF}},b}(t,b^m,q^{m}-1,p(t,\cdot,\cdot)\big) p(t,q^{m}-1,b^m)- \bar \lambda^+ \mathds{1}_{\{ b^m>- \tilde b\}}  p(t,q^{m},b^m-1)\\
&\quad   - \bar \lambda^- \mathds{1}_{\{ b^m< \tilde b\}} p(t,q^{m},b^m+1) - \mathds{1}_{\{ q^{m} < \tilde q^m\}} \Lambda^m \big(\bar \delta^{{\text{\fontsize{4}{4}\selectfont \rm MF}},a}(t,b^m,q^{m}+1,p(t,\cdot,\cdot)) \big) p(t,q^{m}+1,b^m)\\
&\quad+ \bigg(\mathds{1}_{\{ q^{m} < \tilde q^m\}} \Lambda^m  \big(\bar \delta^{{\text{\fontsize{4}{4}\selectfont \rm MF}},b}(t,b^m,q^{m},p(t,\cdot,\cdot))  \big) + \mathds{1}_{\{ q^{m} >- \tilde q^m\}}\Lambda ^m \big(\bar \delta^{{\text{\fontsize{4}{4}\selectfont \rm MF}},a}(t,b^m,q^{m},p(t,\cdot,\cdot)) \big)\\
&\quad + \bar \lambda^- \mathds{1}_{\{ b^m>- \tilde b\}}   +\bar \lambda^+ \mathds{1}_{\{ b^m< \tilde b\}}  \bigg) p(t,q^{m},b^m)=0,\; (t,q^m,b^m)\in (0,T]\times \llbracket -\tilde q^m,\tilde q^m \rrbracket  \times \llbracket - \tilde b , \tilde b \rrbracket,  \\
&p(0,q^{m},b^m) = P_0(q^{m},b^m) ,\; (q^m,b^m)\in  \llbracket -\tilde q^m,\tilde q^m \rrbracket  \times \llbracket - \tilde b , \tilde b \rrbracket  ,
\end{aligned}
\right.
\end{align}
where $P_0$ denotes the probability mass function of the initial inventories of the market-takers and their initial signals at time $0$. Notice that the supremum in the Hamiltonian \eqref{HamMM} is reached at $(d^{b}, d^{a}) := \big( \bar{d}( z - \tilde z^1 ), \bar{d}( z - \tilde z^2 ) \big),$
and the supremum in the Hamiltonian \eqref{HamMT} is reached at $(d^{m,b}, d^{m,a}) := \big( \bar{d}^{m}( z - \tilde z^1  ), \bar{d}^{m}(  z - \tilde z^2 ) \big),$
where $ \bar{d}: p \in \mathbb R \longmapsto \sigma/k + p ,$ and $\bar{d}^{m}: p \in \mathbb R \longmapsto \sigma/k^m + p$. Finally, we arrive to the following master equation for our problem at the equilibrium
\begin{equation}\label{MasterEq}
\left\{\begin{aligned}
0  &=  \partial_t W(t, q) -\phi \sigma^2  q^2\\
 &\quad + q \kappa \int_{\mathbb R^2}   \Lc \Big(y, \bar{d}^{m}\big(W^m(t,z,y)- W^m(t,z,y+1) \big),  \bar{d}^{m}\big(W^m(t,z,y)- W^m(t,z,y-1) \big) \Big) p(t,y,  z)\d y\d z \\
    & \quad + \mathcal{H}\Big(t,q,\bar{d}\big(W(t,q)- W(t,q+1) \big), \bar{d}\big( W(t,q), W(t,q-1) \big),W(t,q),W(t,q+1),W(t,q-1),\bar \delta^{m},p(t,\cdot,\cdot)\Big), \\
  0 &=  \partial_t W^m (t,q^m,b^m) - \gamma \sigma^2 (q^m)^2 + b^mq^m + \Kc\Big(b^m,W^m (t,q^m,b^m),W^m(t,q^m,b^m+1),W^m (t,q^m,b^m-1)\Big)\\
  & \quad+ q^m \kappa \int_{\mathbb R^2}    \Lc \Big(y, \bar{d}^{m}\big(W^m(t,z,y)- W^m(t,z,y+1) \big),  \bar{d}^{m}\big(W^m(t,z,y)- W^m(t,z,y-1) \big) \Big) p(t,y,  z)\d y\d z \\
   &\quad + \mathcal{H}^m \Big(q^m,\bar{d}^{m}\big(W^m(t,q^m,b^m)- W^m(t,q^m+1,b^m) \big),\bar{d}^{m}\big( W^m(t,q^m,b^m), W^m(t,q^m-1,b^m) \big),\\
&   \qquad \qquad \;\;\; W^m(t,q^m,b^m), W^m(t,q^m+1,b^m),W^m(t,q^m-1,b^m)\Big),\\
 0  &=\partial_t p(t,q^{m},b^m) - \mathds{1}_{\{ q^{m}>- \tilde q^m\}} \Lambda^m  \Big( \bar{d}^{m}\big(W^m(t, q^m - 1,b^m)- W^m(t,q^m,b^m) \big)\Big) p(t,q^{m}-1,b^m)\\
&\quad - \mathds{1}_{\{ q^{m} < \tilde q^m\}} \Lambda^m \Big(\bar{d}^{m}\big(W^m(t,q^m+1,b^m)- W^m(t,q^m,b^m) \big) \Big) p(t,q^{m}+1,b^m)\\
&\quad - \bar \lambda^+ \mathds{1}_{\{ b^m>- \tilde b\}}  p(t,q^{m},b^m-1) + \bar \lambda^- \mathds{1}_{\{ b^m< \tilde b\}} p(t,q^{m},b^m+1) \\
&\quad+ \bigg(\mathds{1}_{\{ q^{m} < \tilde q^m\}} \Lambda^m  \Big(\bar{d}^{m}\big(W^m(t, q^m,b^m )- W^m(t,q^m+1,b^m) \big)  \Big) \\
&\quad + \mathds{1}_{\{ q^{m} >- \tilde q^m\}}\Lambda ^m \Big(\bar{d}^{m}\big(W^m(t, q^m,b^m )- W^m(t,q^m-1,b^m) \big) \Big) + \bar \lambda^- \mathds{1}_{\{ b^m>- \tilde b\}}   +\bar \lambda^+ \mathds{1}_{\{ b^m< \tilde b\}}   \bigg) p(t,q^{m},b^m),
\end{aligned}    
\right.
\end{equation}
where $\bar \delta^m = \big(\bar \delta^{m,b}, \bar \delta^{m,a} \big) \in (\Dc^m)^2$ is the function such that 
\begin{align*}
\bar \delta^{m,b}(t,b^m,q^m,p(t,\cdot,\cdot)) &= \bar{d}^{m}\big(W^m(t,q^m,b^m)- W^m(t,q^m+1,b^m) \big),\\
\bar \delta^{m,a}(t,b^m,q^m,p(t,\cdot,\cdot)) &= \bar{d}^{m}\big(W^m(t,q^m,b^m)- W^m(t,q^m-1,b^m) \big),
\end{align*}
and where $p$ is the flow of probability mass functions associated with $\Pi$, with initial and terminal conditions
\begin{align}\label{termcondme}
p(0,q^{m},b^m) = P_0(q^{m},b^m) , \; W(T,q) = W^m (T,q^m,b^m) = 0,\; \text{for $(b^m,q^m, q) \in \llbracket - \tilde b , \tilde b \rrbracket \times \llbracket - \tilde q^m , \tilde q^m \rrbracket \times \llbracket - \tilde q , \tilde q \rrbracket$.}
\end{align}

\endgroup
\subsection{Main results}
The following result guarantees existence of a solution to \eqref{MasterEq}, its proof is relegated to \Cref{sec:existence}. For simplicity of notations, let us introduce in what follows $\mathcal B := \llbracket - \tilde b , \tilde b \rrbracket$, $\mathcal Q^m := \llbracket -\tilde q^m, \tilde q^m \rrbracket$,   $\mathcal Q := \llbracket -\tilde q, \tilde q \rrbracket$. 
\begin{theorem}\label{existence}
The master equation \eqref{MasterEq} with terminal conditions \eqref{termcondme} admits a solution $\big(W, W^m, p\big)$, where $W$ is $C^1$ in time and bounded on $[0,T] \times \mathcal Q,$ $W^m$ is $C^1$ in time and bounded on $[0,T]  \times \mathcal Q^m\times \mathcal B,$ and $p$ is $C^1$ in time and bounded on $[0,T]  \times \mathcal Q^m \times \mathcal B$. 
\end{theorem}

We conclude this section with a verification theorem, whose proof can be found in \Cref{sec:verif}.
 
\begin{theorem}\label{verif}
Let $\tilde \Pi  \in C \big([0,T], \mathcal P(\mathcal Q^m\times \mathcal B) \big)$ be a flow of probability measures on $\mathcal Q^m\times  \mathcal B$. We denote by $\tilde p $ the associated probability mass functions. Assume that the coupled Hamilton--Jacobi--Bellman equation
\begin{align*}
\left\{\begin{aligned}
0  &=  \partial_t \tilde W(t, q) -\phi \sigma^2  q^2\\
 &\quad + q \kappa \int_{\mathbb R^2}   \Lc \Big(y, \bar{d}^{m}\big(\tilde W^m(t,z,y)- \tilde W^m(t,z,y+1) \big),  \bar{d}^{m}\big(\tilde W^m(t,z,y)- \tilde W^m(t,z,y-1) \big) \Big) \tilde p(t,y,  z)\d y\d z\\
   &  \quad + \mathcal{H}\Big(t,q,\bar{d}\big(\tilde W(t,q)- \tilde W(t,q+1) \big), \bar{d}\big( \tilde W(t,q), \tilde W(t,q-1) \big),\tilde W(t,q),\tilde W(t,q+1),\tilde W(t,q-1), \bar \delta^m, \tilde p(t,\cdot,\cdot)\Big), \\
  0 &=  \partial_t \tilde W^m (t,q^m,b^m) - \gamma \sigma^2 (q^m)^2 + b^mq^m + \Kc\Big(b^m,\tilde W^m (t,q^m,b^m),\tilde W^m(t,q^m,b^m+1),\tilde W^m (t,q^m,b^m-1)\Big)\\
  & \quad+ q^m \kappa \int_{\mathbb R^2}    \Lc \Big(y, \bar{d}^{m}\big(\tilde W^m(t,z,y)- \tilde W^m(t,z,y+1) \big),  \bar{d}^{m}\big(\tilde W^m(t,z,y)- \tilde W^m(t,z,y-1) \big) \Big) \tilde p(t,y,  z)\d y\d z\\
 &  \quad + \mathcal{H}^m \Big(q^m,\bar{d}^{m}\big(\tilde W^m(t,q^m,b^m)- \tilde W^m(t,q^m+1,b^m) \big),\bar{d}^{m}\big( \tilde W^m(t,q^m,b^m), \tilde W^m(t,q^m-1,b^m) \big),\\
  & \qquad \qquad  \;\;\tilde W^m(t,q^m,b^m), \tilde W^m(t,q^m+1,b^m),\tilde W^m(t,q^m-1,b^m)\Big),
\end{aligned}\right.    
\end{align*}
with terminal condition $\tilde W(T,.) = \tilde W^m(T,\cdot,\cdot) = 0 $ has a classical solution $(\tilde W, \tilde W^m)$ once continuously differentiable in time. We introduce the following feedback controls 
\begin{align}
\label{cloptcont0bcor}
\begin{split}
      \bar \delta^{\star,b}(t,q, \tilde p(t,\cdot,\cdot)) = \frac{\sigma}{k}+ \tilde W (t,q)-&\tilde W (t,q+1),\; \bar \delta^{\star,a}(t,q, \tilde p(t,\cdot,\cdot)) = \frac{\sigma}{k}+ \tilde W (t,q)-\tilde W (t,q-1), \\
     \bar \delta^{m,\star, b} (t,b, q^m, \tilde p(t,\cdot,\cdot)&) =  \frac{\sigma}{k^m} + \tilde W^m(t, q^m, b) - \tilde W^m(t, q^m +1, b), \\
     \bar \delta^{m,\star, a} (t,b, q^m, \tilde p(t,\cdot,\cdot)&) =  \frac{\sigma}{k^m} + \tilde W^m(t, q^m, b) - \tilde W^m(t, q^m -1, b) .
     \end{split}
 \end{align}
Then the functions $\tilde W$ and $\tilde W^m$ correspond respectively to the value functions of the market-maker and the market-takers associated with the flow of probability measures $\tilde \Pi,$ in the sense that
\[\tilde W (0, q_0) =  \tilde V (\bar \delta^m, \tilde \Pi),\; \text{\rm and}\; \tilde W^m (0, q^m_0, b^m_0) =  \tilde V^m (\bar \delta^m, \tilde \Pi),
\]
and the controls given in closed-loop by \eqref{cloptcont0bcor} are optimal.
\end{theorem}

From these two results , we immediately get the following corollary, which provides us with a Markovian equilibrium.
\begin{corollary}\label{corver}
Consider the solution $\big(W, W^m, p\big)$ to the master equation \eqref{MasterEq} with terminal conditions \eqref{termcondme} defined in {\rm \Cref{existence}}. We introduce the associated feedback controls as in {\rm\Cref{verif}}
\begin{equation}
\label{cloptcont0b}
\begin{split}
      \bar \delta^{\star,b}(t,q, p(t,\cdot,\cdot)) = \frac{\sigma}{k}+  W (t,q)- &W (t,q+1),\; \bar \delta^{\star,a}(t,q,  p(t,\cdot,\cdot)) = \frac{\sigma}{k}+  W (t,q)-\tilde W (t,q-1), \\
      \bar \delta^{m,\star, b} (t,b, q^m,  p(t,\cdot,\cdot)&) =  \frac{\sigma}{k^m} + W^m(t, q^m, b) - W^m(t, q^m +1, b), \\
      \bar \delta^{m,\star, a} (t,b, q^m, p(t,\cdot,\cdot)&) =  \frac{\sigma}{k^m} +  W^m(t, q^m, b) -W^m(t, q^m -1, b) .
 \end{split}
 \end{equation}
These feedback controls and the flow $\Pi \in \mathcal D^p$ constitute a Markovian equilibrium in the sense of {\rm \Cref{markoveq}}.
\end{corollary}

\begin{remark}
{\rm\Cref{corver}} gives us an equilibrium of our mean-field game. The presence of controlled point processes in the problem makes the uniqueness of the equilibrium hard to prove, as classical techniques do not apply here. This matter is left for a future, more general work. However, the absence of numerical instability in our numerous experiments leads us to believe that the equilibrium is indeed unique.
\end{remark}
\section{Numerical results}\label{numsec}

In this section, we apply our model to the case of an asset with the characteristics in \Cref{tab}. This time horizon ensures convergence towards stationary quotes at time $t=0$. We consider that the asset is traded by a constant trade size of $10$ assets. We solve the master equation \eqref{MasterEq} using three explicit Euler schemes, with a loop in order to find the fixed-point defined in \Cref{existence}.
\begin{table}[H]\label{tab}
\small
\begin{center}
\begin{tabular}{|c | c|} 
 \hline 
Parameter & Value \\ [0.5ex] 
 \hline
 Daily volatility & $\sigma = 8\  \textrm{\$} \cdot \textrm{day}^{-\frac{1}{2}}$ \\ [0.5ex] 
 Permanent market impact & $\kappa = 5 \cdot 10^{-5}\ \textrm{\$}$ \\ [0.5ex]
 Intensity parameters for the signals & $\bar \lambda^+ = \bar \lambda^- = 1.5\ \textrm{\$}\cdot \textrm{day}^{-2}$ \\ [0.5ex] 
Signal limit & $\tilde b = 4\ \textrm{\$}\cdot \textrm{day}^{-1}$ \\ [0.5ex] 
Intensity function of the market-takers & $A^m =70\ \textrm{day}^{-1}$ and $k^m=19\ \textrm{day}^{-\frac{1}{2}}$ \\ [0.5ex] 
Intensity function of the market-maker & $A =100\ \textrm{day}^{-1}$ and $k=19\ \textrm{day}^{-\frac{1}{2}}$ \\ [0.5ex] 
 Risk limits & $\tilde q  = \tilde q^m = 120$ \\ [0.5ex] 
 Risk aversion of the market-takers & $\gamma = 1\cdot10^{-4}\ \textrm{\$}^{-1}$ \\ [0.5ex]
 Risk aversion of the market-maker & $\phi = 5\cdot10^{-4}\ \textrm{\$}^{-1}$ \\ [0.5ex] 
 Time horizon & $T = 5\ \textrm{days}$ \\ 
 \hline 
\end{tabular}
\end{center}

\vspace{-1em}
\caption {\small Value of the parameters.}
\label{table:1DParams}
\end{table}

\subsection{Market-takers' behaviour: probability mass function}

\begin{figure}[!ht]\centering
\includegraphics[width=0.64\textwidth]{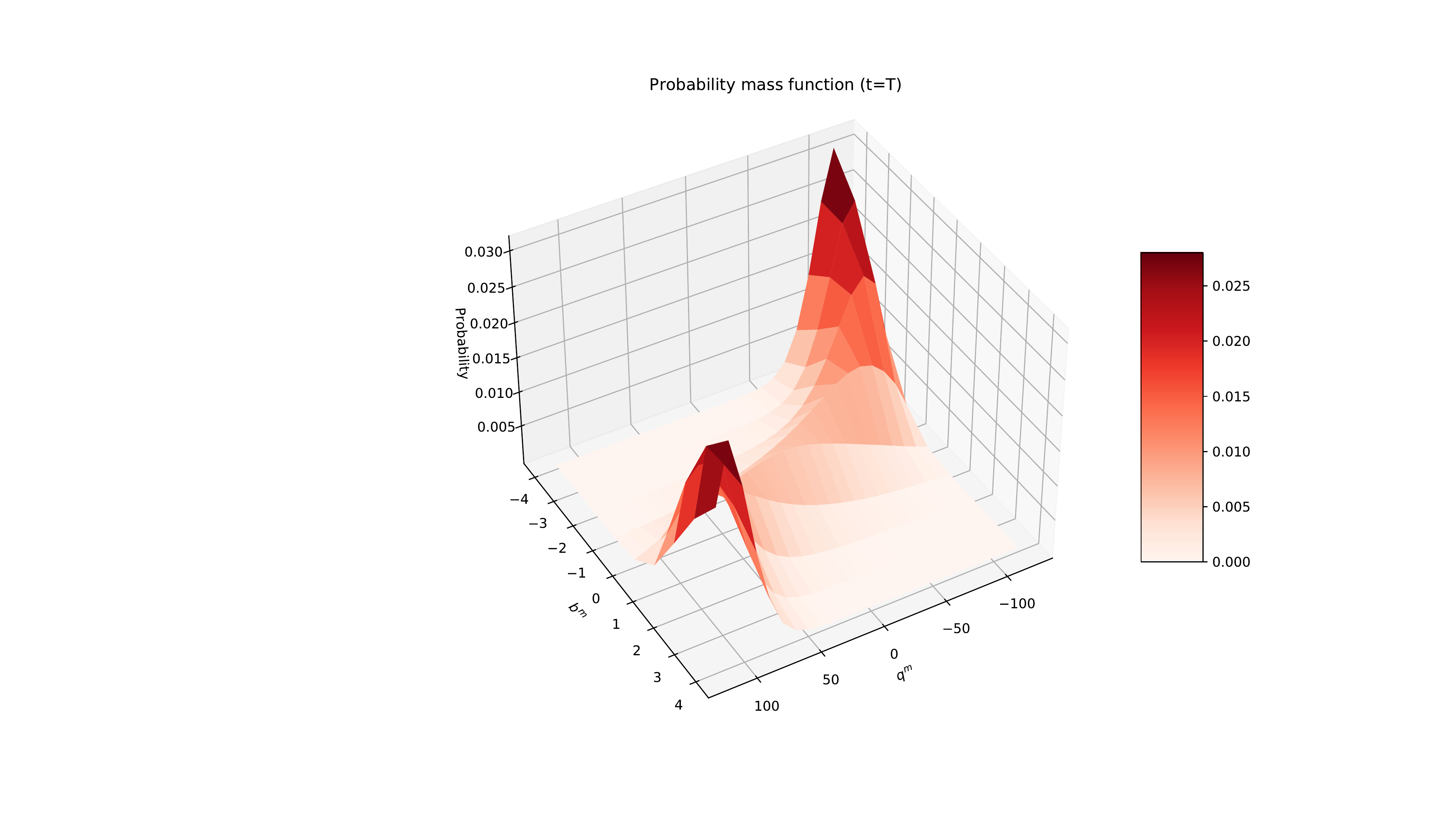}

\vspace{-2em}
\caption{\small Joint distribution of $b^m$ and $q^m$.}\label{p_mass_3D}
\end{figure}

\begin{figure}[!ht]
\centering
\hspace*{-2.5cm}
\begin{subfigure}{.8\textwidth}
  \centering
  \includegraphics[width=.7\linewidth]{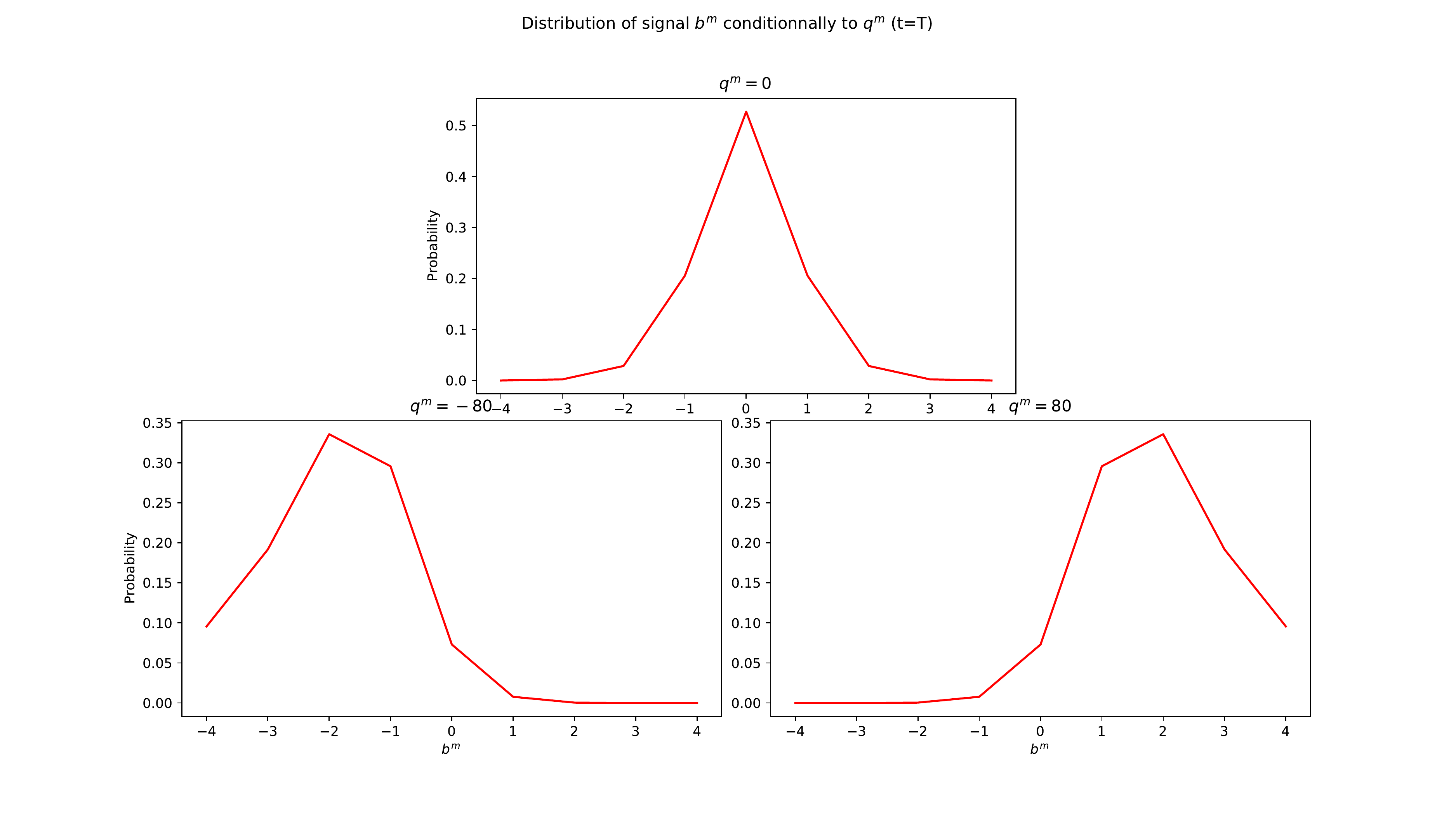}
  \vspace{-1.5em}
  \caption{\small Distribution of $b^m$ for a fixed $q^m$.}
  \label{p_mass_b}
\end{subfigure}%
\hspace*{-6cm}
\begin{subfigure}{.8\textwidth}
  \centering
  \includegraphics[width=.7\linewidth]{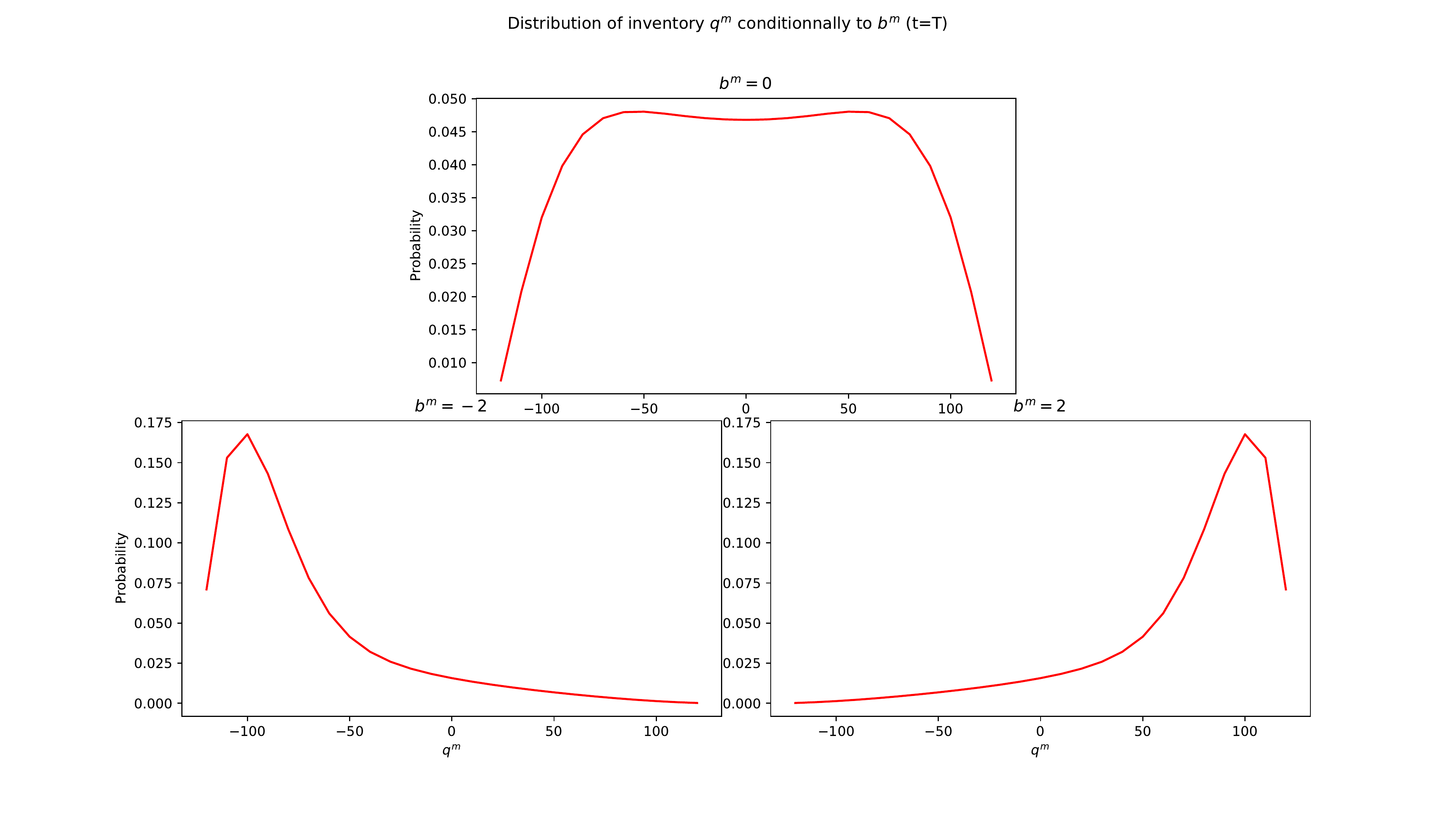}
  \vspace{-1.5em}
  \caption{\small Distribution of $q^m$ for a fixed $b^m$.}
  \label{p_mass_q}
\end{subfigure}
\caption{\small Conditional distributions of $b^m$ and $q^m$.}
\label{p_mass}
\end{figure}

The probability mass function $p$ is plotted at time $t=T$ in \Cref{p_mass_3D}. We then plot in \Cref{p_mass_b} the distribution of the signal $b^m$ conditionally to the value of the inventory $q^m$ (for $q^m = 0$, $q^m = -80$, and $q^m = 80$). As expected, the distribution is symmetric when $q^m = 0$. Moreover, when the inventory is very negative, the probability of having a positive signal is almost $0$ and the probability of having a negative signal is high, and conversely for a very positive inventory. We finally plot in \Cref{p_mass_q} the distribution of the inventory $q^m$ conditionally to the value of the signal $b^m ($for $b^m = 0$, $b^m = -2$, and $b^m = 2$), and observe the same kind of expected behaviour :  the distribution is symmetric when $b^m = 0$, whereas all the weight is on the negative (resp. positive) values of $q^m$ when $b^m = -2$ (resp. $b^m = 2$). 

\vspace{-1.5em}

\subsection{Market-takers' behaviour: value function and optimal quotes}

\begin{figure}[!ht]\centering
\includegraphics[width=0.7\textwidth]{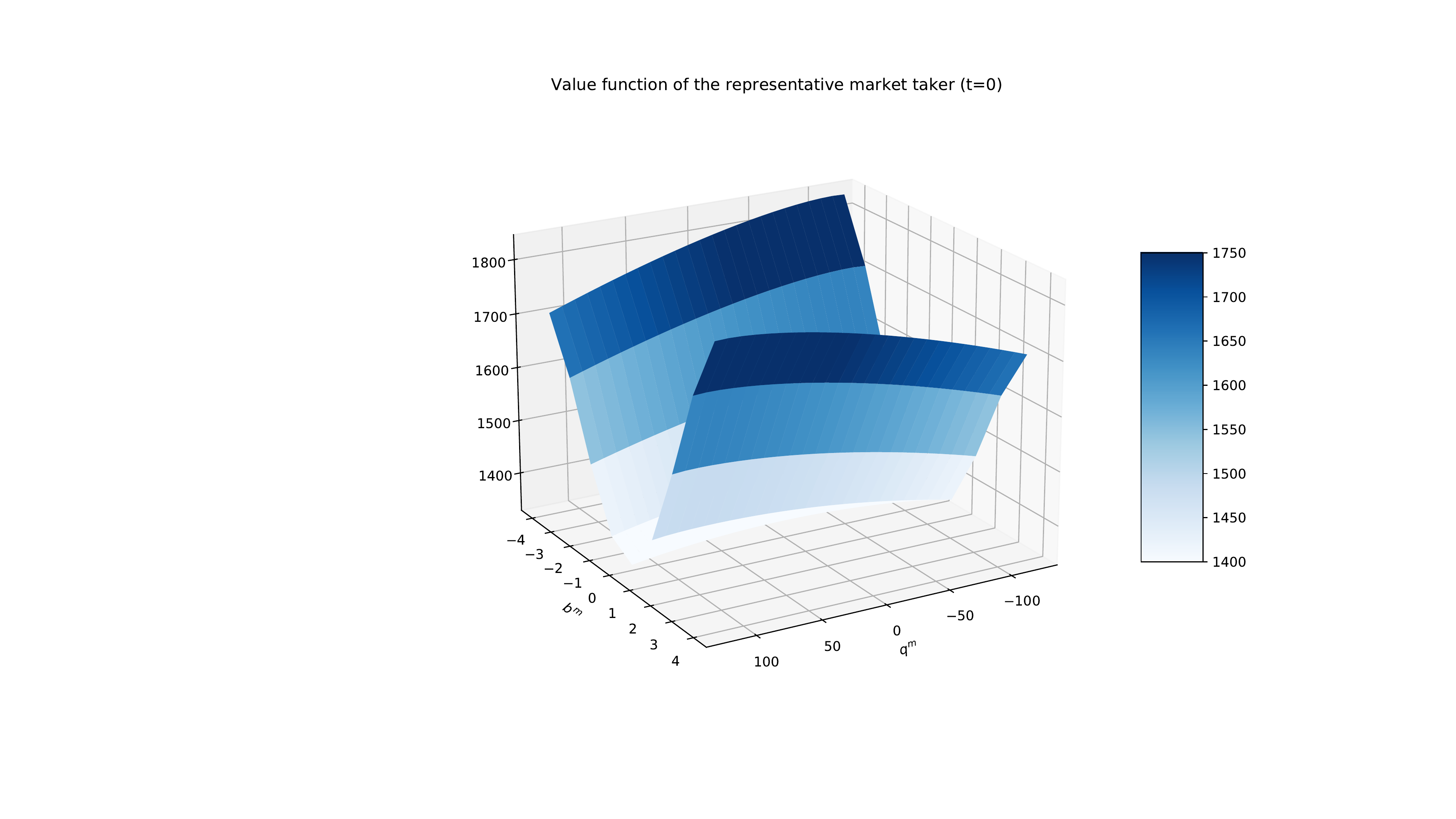}

\vspace{-2em}
\caption{\small Value function of the representative market-taker.}\label{V_minor_3D}
\end{figure}

\begin{figure}[!ht]
\centering
\hspace*{-2.5cm}
\begin{subfigure}{.8\textwidth}
  \centering
  \includegraphics[width=.7\linewidth]{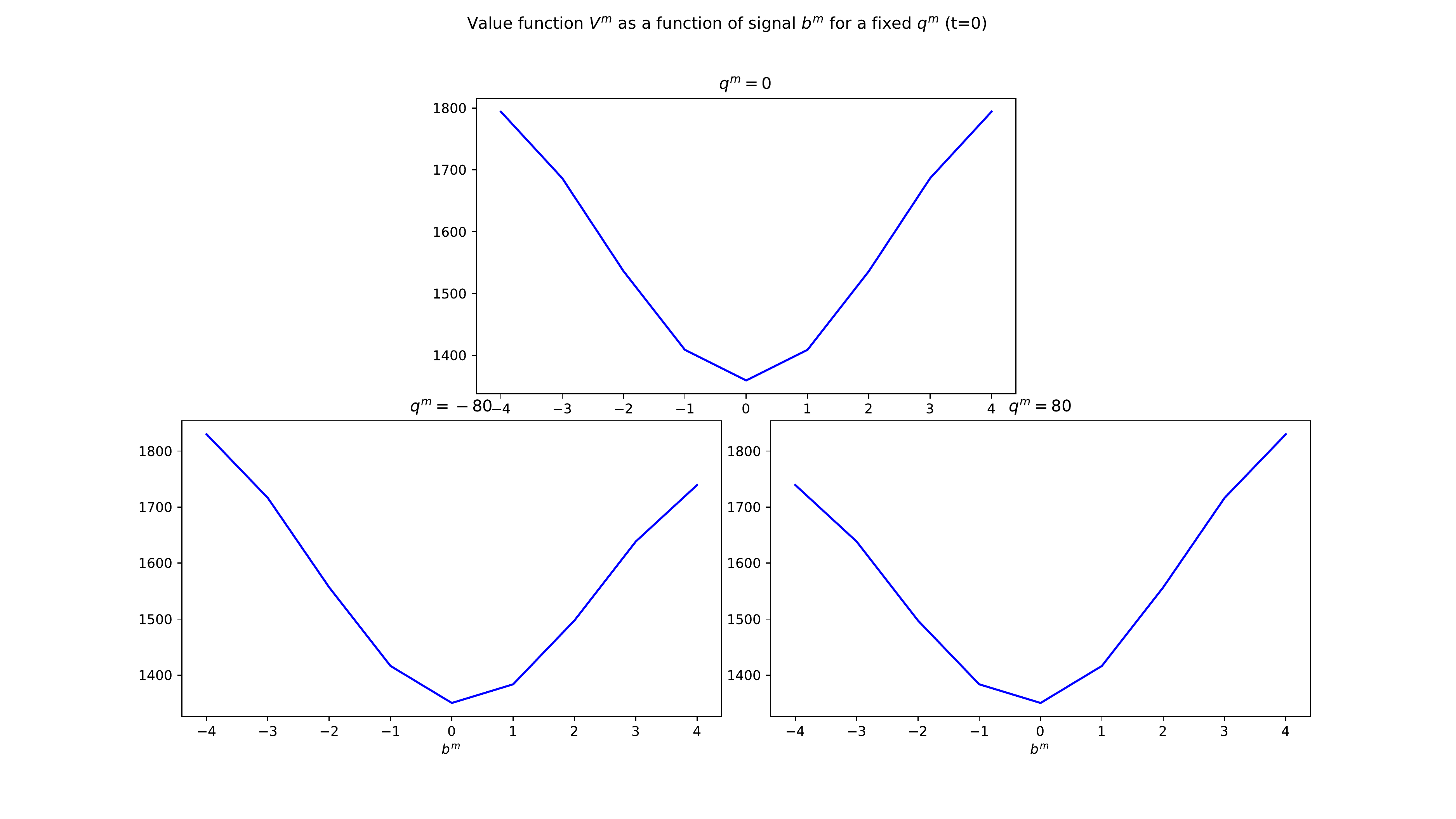}
  \vspace{-2em}
  \caption{\small For a fixed $q^m$.}
  \label{V_minor_b}
\end{subfigure}%
\hspace*{-6cm}
\begin{subfigure}{.8\textwidth}
  \centering
  \includegraphics[width=.7\linewidth]{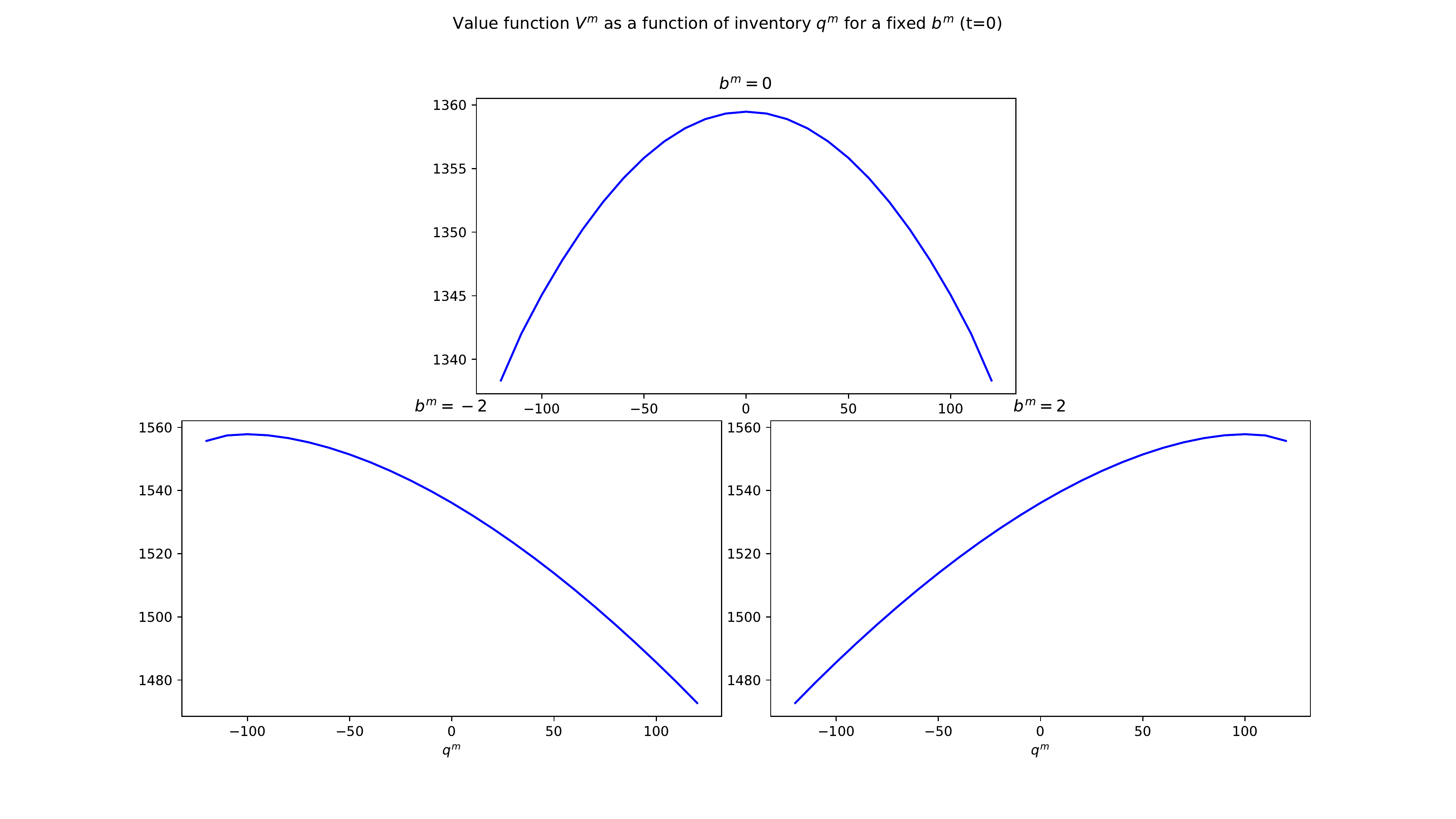}
  \vspace{-2em}
  \caption{\small For a fixed $b^m$.}
  \label{V_minor_q}
\end{subfigure}
\caption{Value function of the representative market-taker.}
\label{V_minor_slices}
\end{figure}

The value function $W^m$ is plotted at time $t=0$ in \Cref{V_minor_3D}. We then plot in \Cref{V_minor_b} the same value function at fixed values of $q^m$, and observe some interesting behaviours. Indeed, when $q^m=0$, the value function is symmetric but its minimum value is in $b^m = 0$, meaning that even with an empty inventory, the market-taker always prefer having a positive or negative signal because this will allow him to make more profit over the time period $[0,T]$. This effect is even more striking in the cases $q^m = -80$ and $q^m = 80$: even with an already very long or short inventory, the worst case for the market-taker is a flat signal. Indeed, when the market-taker is short, he prefers having a negative signal, but a positive signal is still better than a flat signal because he knows she can trade fast enough to become long before the signal changes. Finally, we plot in \Cref{V_minor_b} the value function at fixed values of $b^m$. The results are less surprising and match our intuition: when the signal is flat, the value function is symmetric and its maximum is in $0$, because the market-taker does not have any interest in deviating from an empty position. When the signal is negative, the optimum of the value function is in the negative values for the inventory, and conversely.

\vspace{-1em}

\begin{figure}[!ht]\centering
\includegraphics[width=0.8\textwidth]{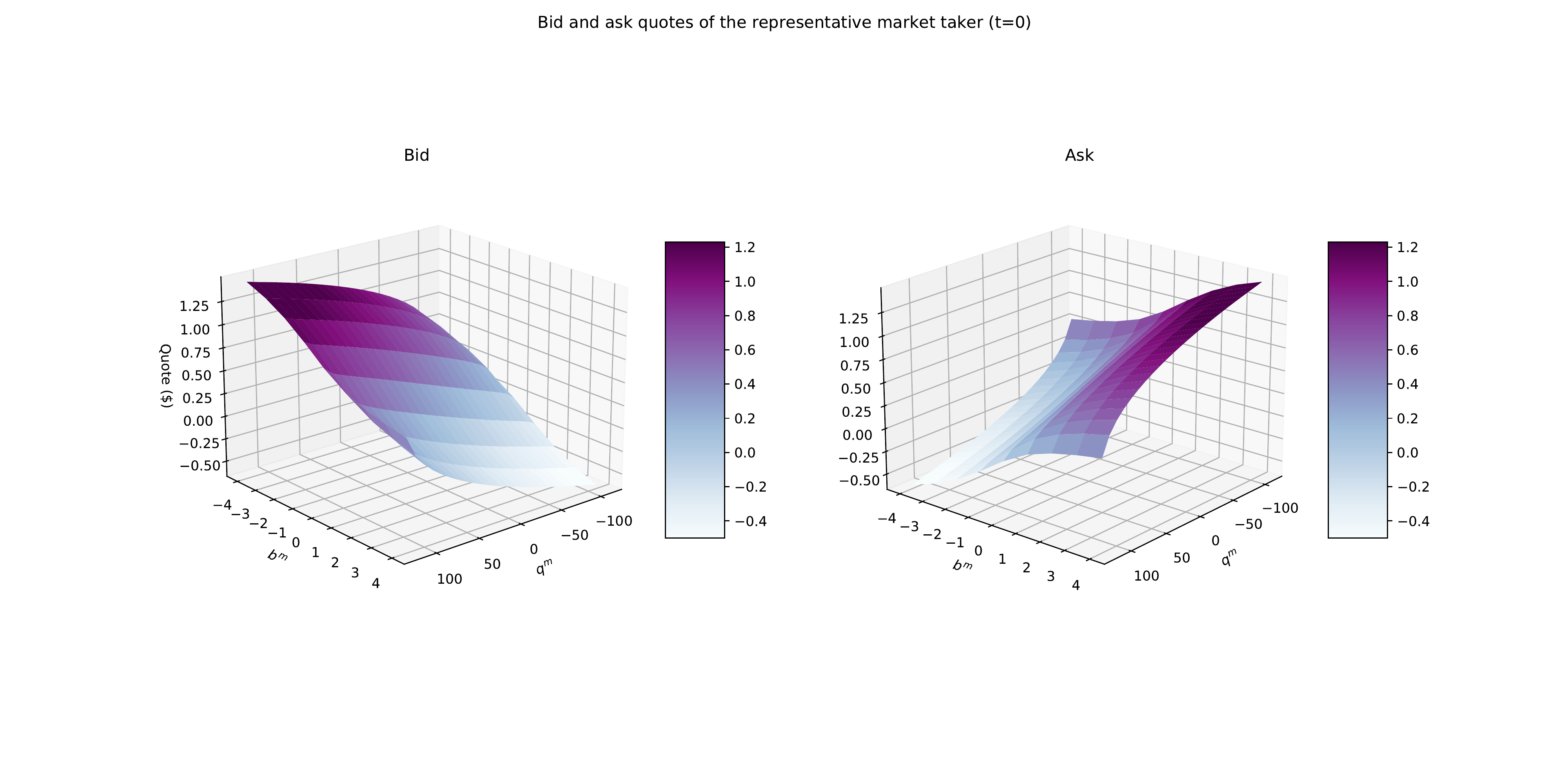}

\vspace{-4em}
\caption{\small Optimal bid and ask quotes of the representative market-taker.}\label{Quotes_minor_3D}
\end{figure}

\begin{figure}[!ht]
\centering
\hspace*{-2.5cm}
\begin{subfigure}{.8\textwidth}
  \centering
  \includegraphics[width=.7\linewidth]{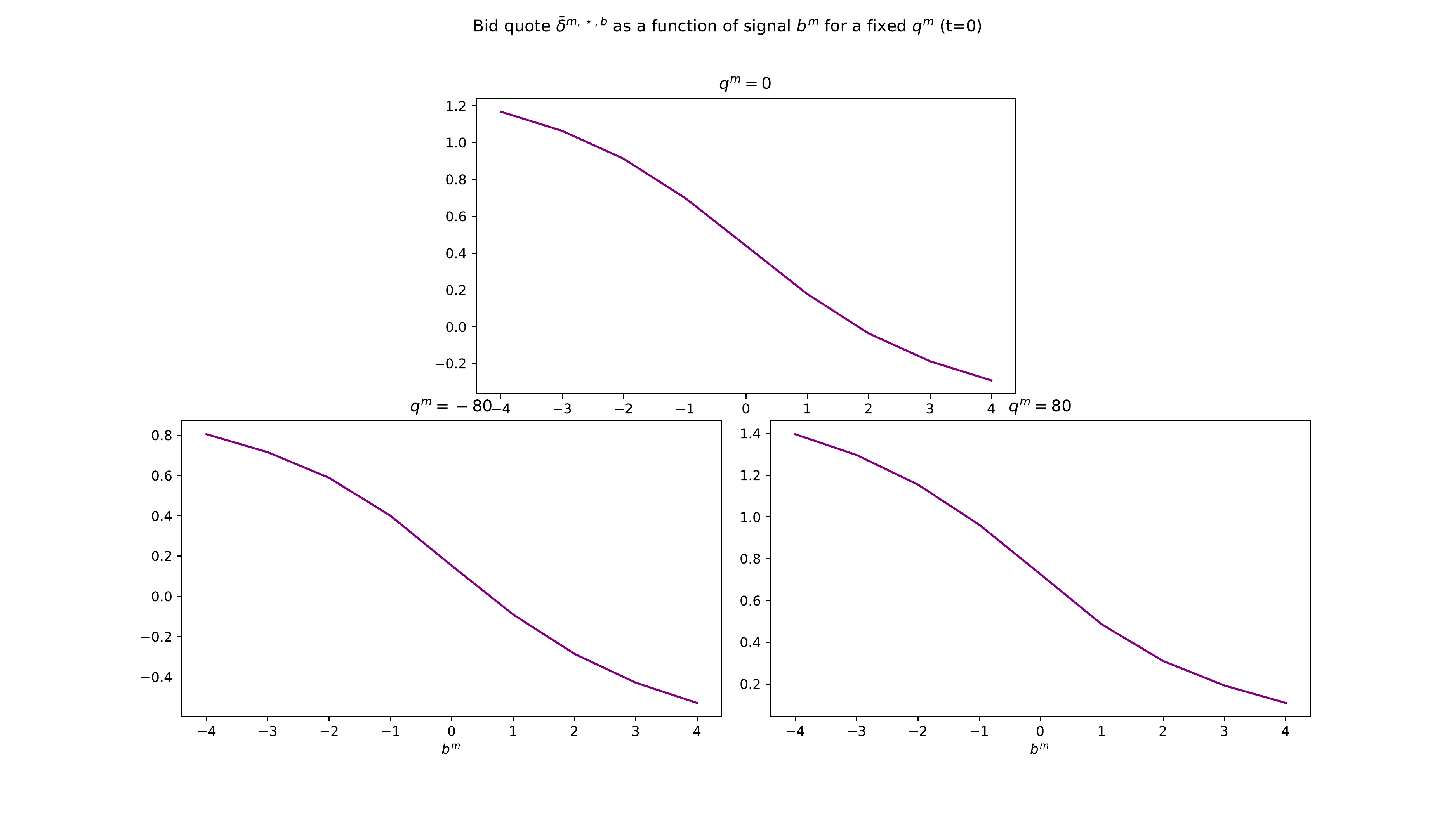}
  \vspace{-2em}
  \caption{\small For a fixed $q^m$.}
  \label{Bid_minor_b}
\end{subfigure}%
\hspace*{-6cm}
\begin{subfigure}{.8\textwidth}
  \centering
  \includegraphics[width=.7\linewidth]{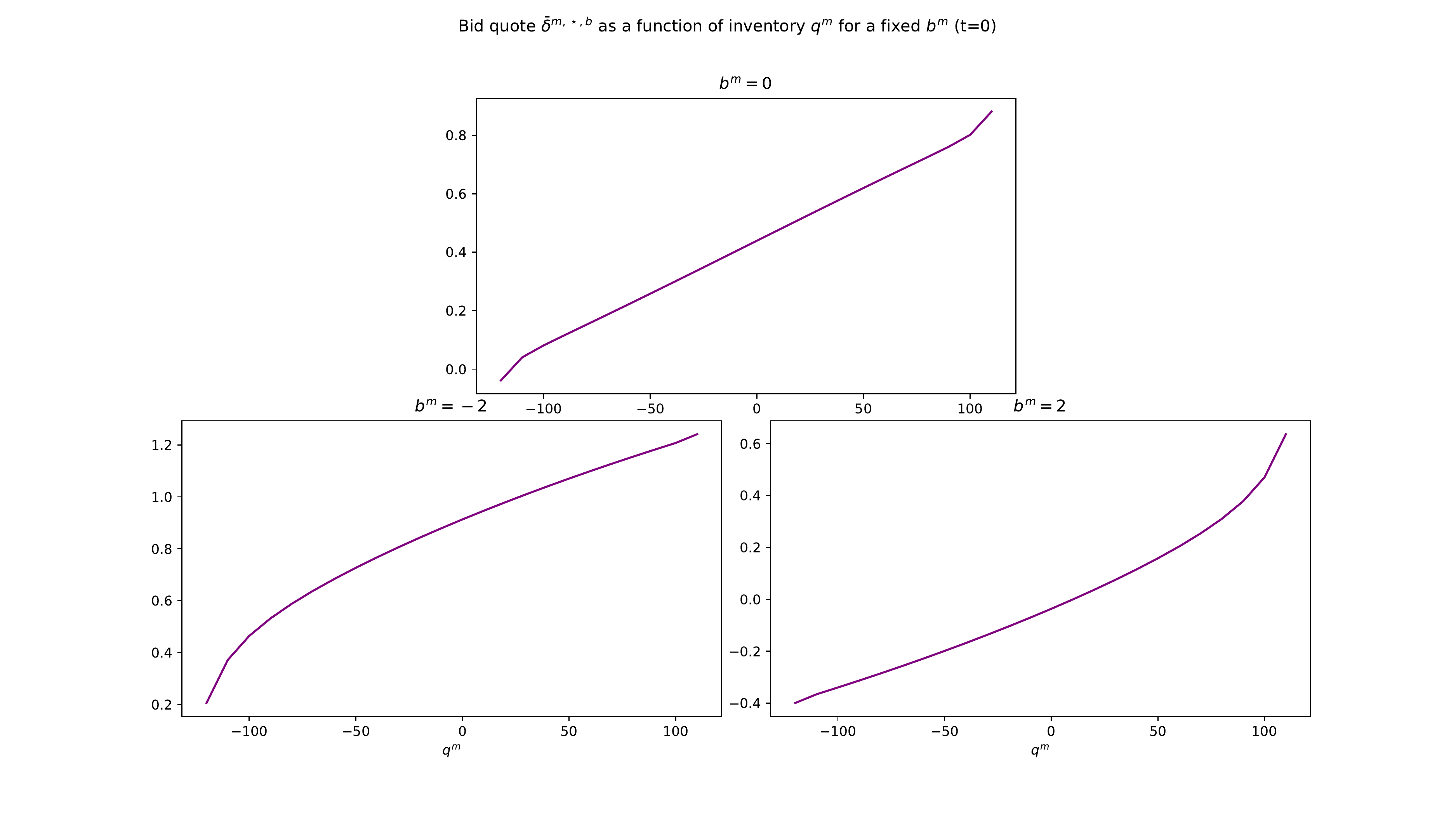}
  \vspace{-2em}
  \caption{\small For a fixed $b^m$.}
  \label{Bid_minor_q}
\end{subfigure}
\caption{\small Optimal bid quote of the representative market-taker.}
\label{Bid_minor}
\end{figure}

We plot the (stationary) optimal bid and ask quotes of the market-taker as a function of $q^m$ and $b^m$ in \Cref{Quotes_minor_3D}. In \Cref{Bid_minor_b}, we plot his bid quotes as a function of $b^m$ for different values of $q^m.$ We observe that when his signal is positive enough, if his inventory is non-positive, he agrees to buy the asset at a price higher than the mid price. In particular when $q^m = -80$, this is true as soon as $b^m$ is positive. Finally in \Cref{Bid_minor_q}, we plot the optimal bid quotes as a function $q^m$ for different values of $b^m$. Again, the result is expected: when the signal is negative, the bid quotes is always high in comparison to the other cases. Conversely, for a positive signal, if his inventory is negative, the market-maker quotes a negative mid-to-bid, meaning that he agrees to buy at a price higher than the mid price.

\subsection{Market-maker's behaviour: value function and optimal quotes}

\begin{figure}[!ht]\centering
\includegraphics[width=0.65\textwidth]{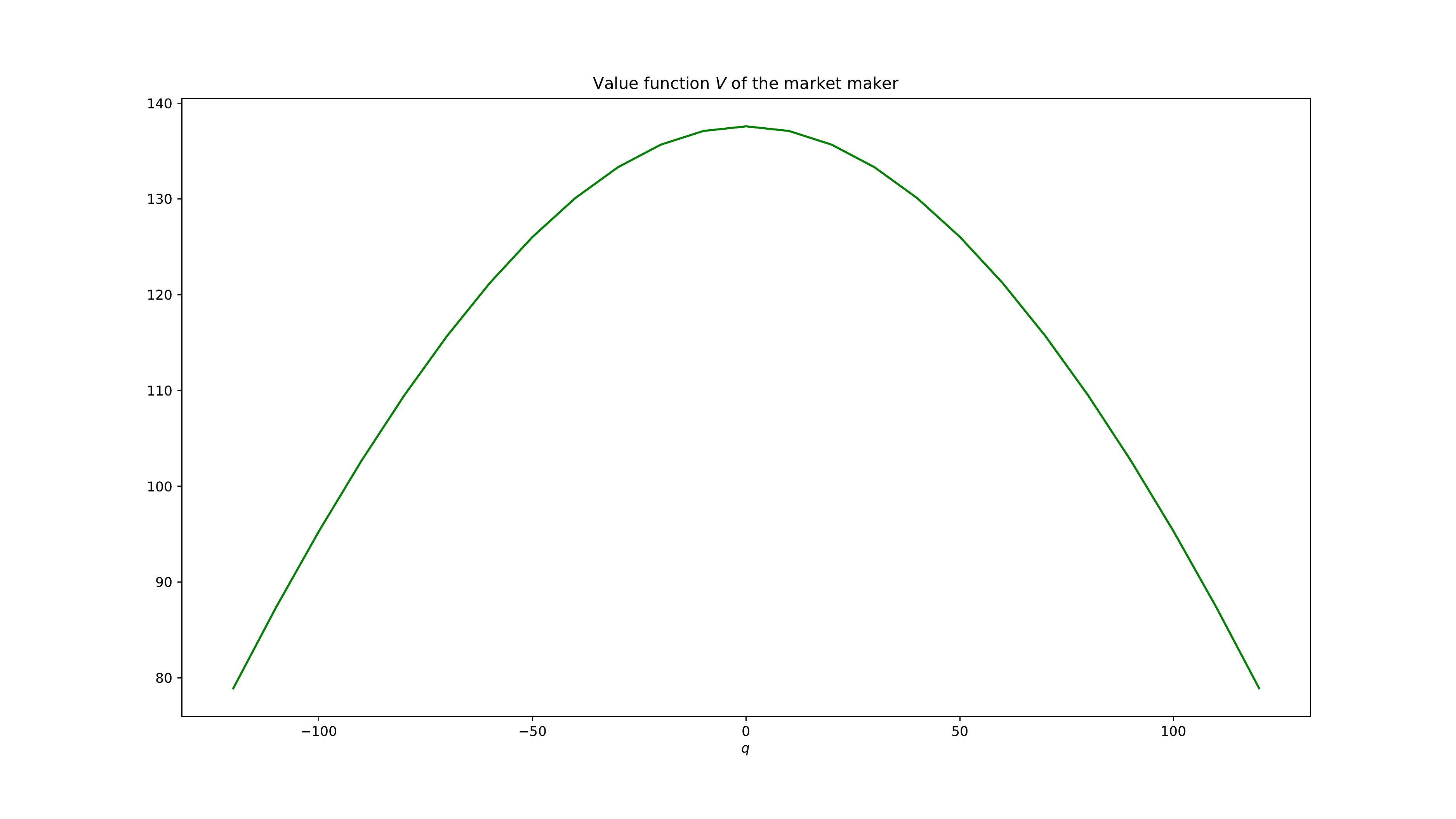}

\vspace{-2em}
\caption{\small Value function of the market-maker.}\label{V_major}
\end{figure}

\begin{figure}[!ht]\centering
\includegraphics[width=0.6\textwidth]{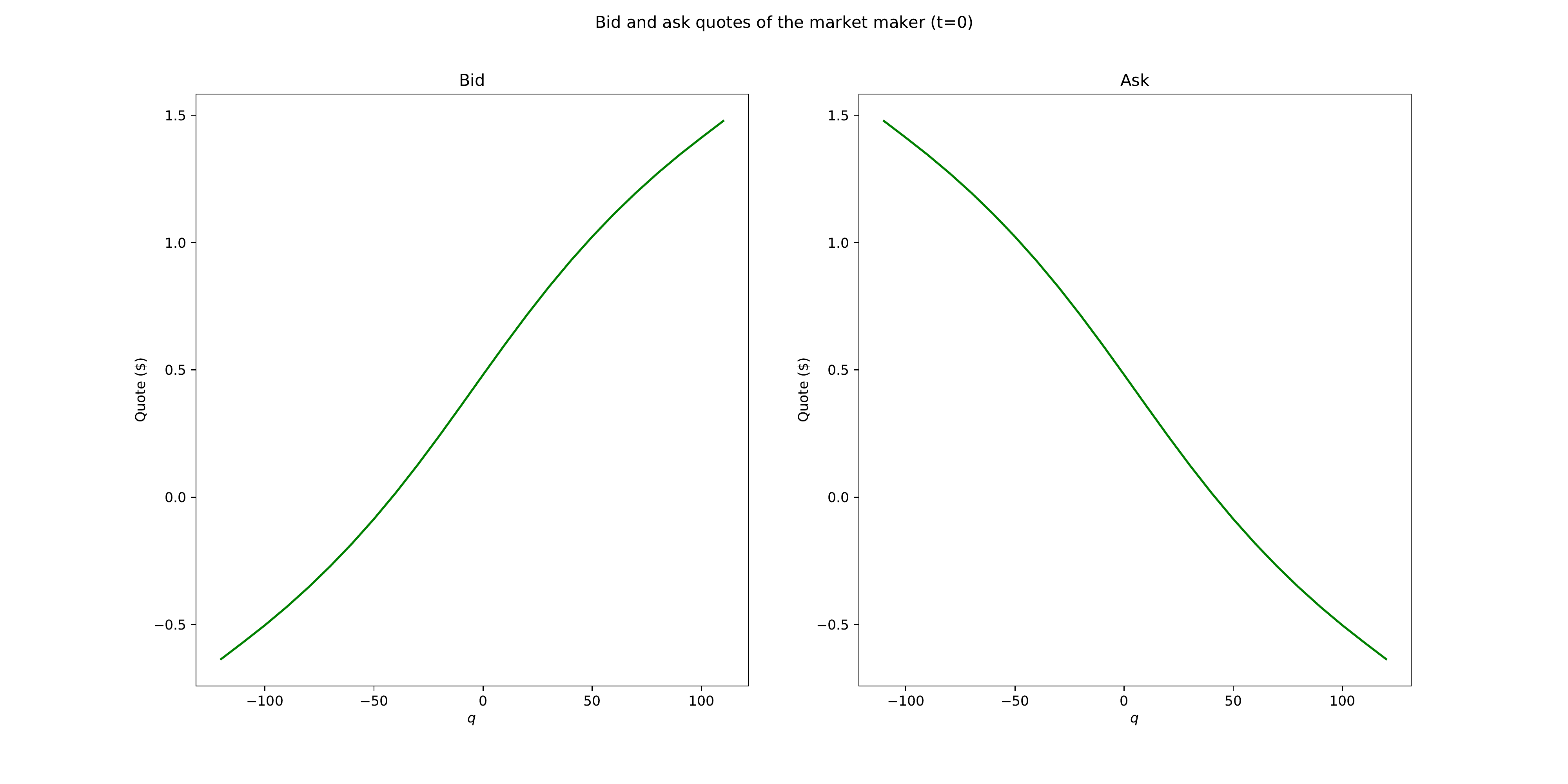}

\vspace{-2em}
\caption{\small Optimal bid and ask quotes of the market-maker.}\label{Quotes_major}
\end{figure}

We plot in \Cref{V_major} the value function of the market-maker as a function of her inventory. As expected, as the market-maker is not reacting to any signal, but only make profit by providing liquidity while managing her inventory risk, the value function is symmetric with a maximum at 0. The optimal bid and ask quotes as functions of her inventory are plotted in \Cref{Quotes_major}, with the expected behaviour: the bid quote increases with the inventory, while the ask quote decreases. The quotes of the market-maker go higher than that of the market-taker because she is more risk-averse.

\subsection{Impact of a more volatile signal }

We now multiply by 10 the intensities $\lambda^+$and $\lambda^-$, in order to study what happens when the trading rate of the market-taker is low compared to the rate at which the signal changes.
 
\begin{figure}[!ht]\centering
\includegraphics[width=0.65\textwidth]{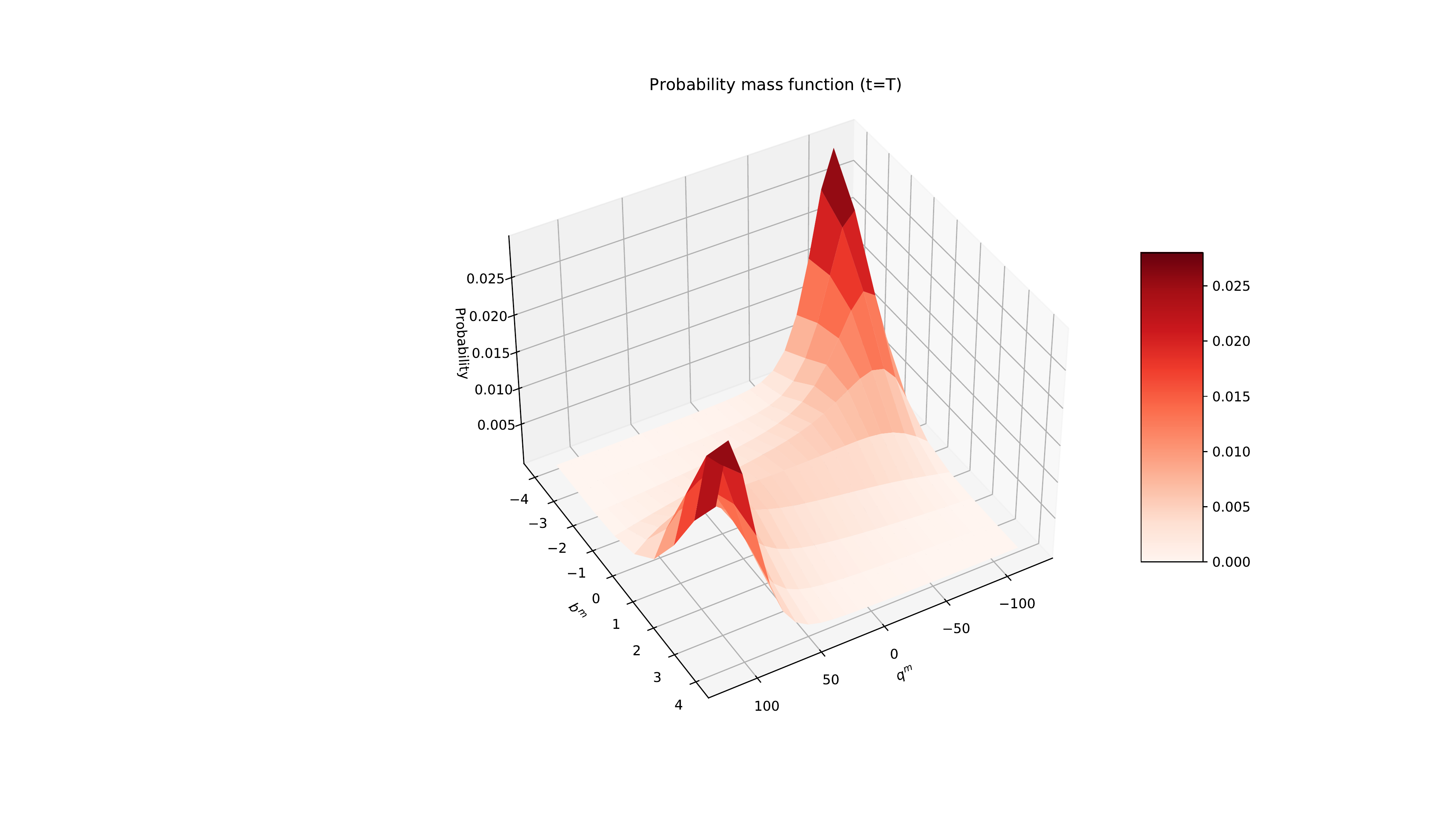}

\vspace{-2em}
\caption{\small Joint distribution of $b^m$ and $q^m$.}\label{p_mass_3D_TR}
\end{figure}

\begin{figure}[!ht]
\centering
\hspace*{-2.5cm}
\begin{subfigure}{.8\textwidth}
  \centering
  \includegraphics[width=.7\linewidth]{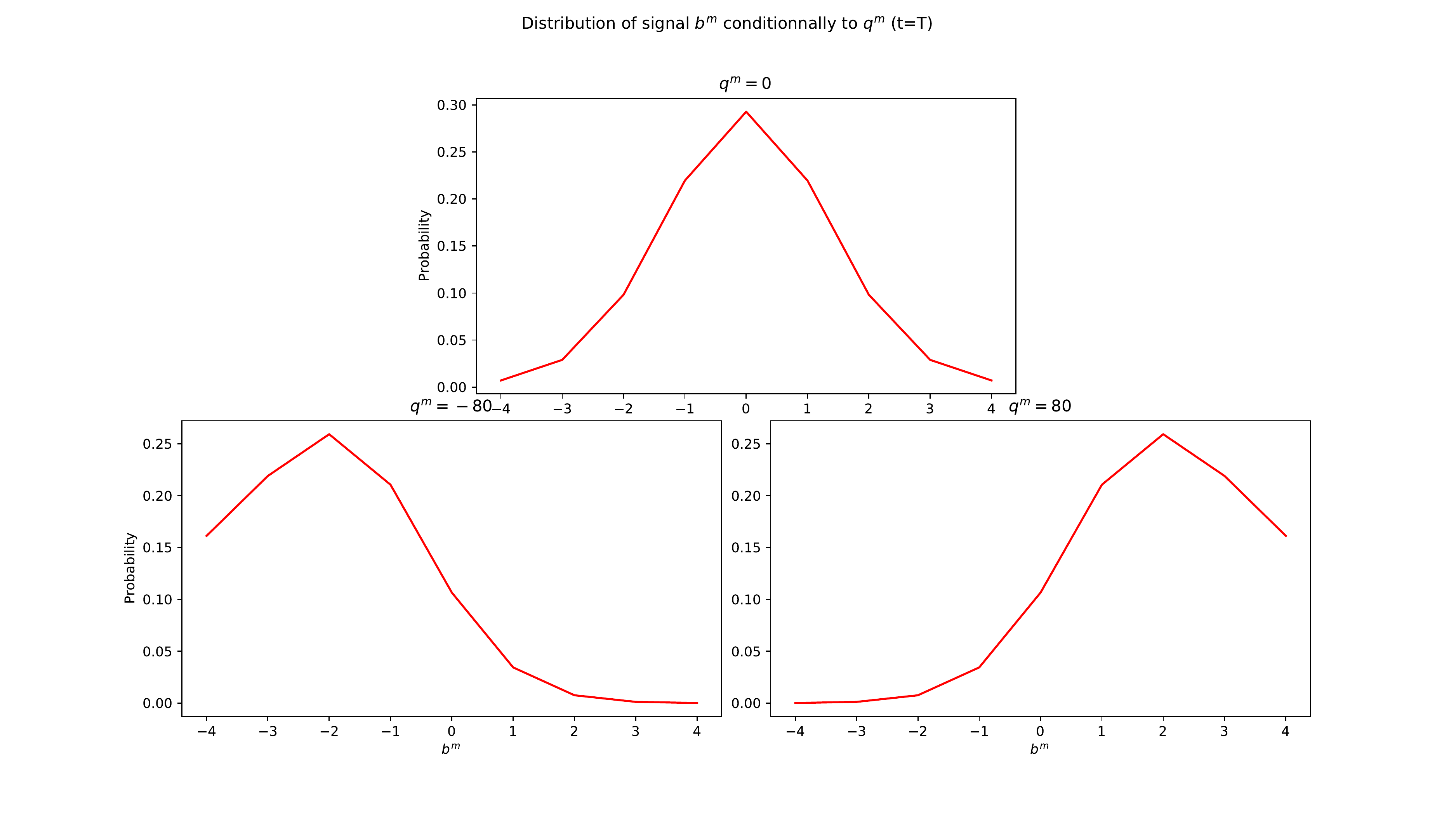}
  \vspace{-2em}
  \caption{\small Distribution of $b^m$ for a fixed $q^m$.}
  \label{p_mass_b_TR}
\end{subfigure}%
\hspace*{-6cm}
\begin{subfigure}{.8\textwidth}
  \centering
  \includegraphics[width=.7\linewidth]{images/p_mass_q_TR.pdf}
  \vspace{-2em}
  \caption{\small Distribution of $q^m$ for a fixed $b^m$.}
  \label{p_mass_q_TR}
\end{subfigure}
\caption{Conditional distributions of $b^m$ and $q^m$.}
\label{p_mass_TR}
\end{figure}

\medskip
As before, we plot in \Cref{p_mass_3D_TR}, \Cref{p_mass_b_TR}, and \Cref{p_mass_q_TR} the probability mass function $p$. We notice in particular the difference between \Cref{p_mass_q} and \Cref{p_mass_q_TR}: while the distribution of inventories is still symmetric around $0$ when the signal is 0, it is now made of two `bumps'. This is because traders cannot react fast enough to a change of signal, so they cannot always get back to an empty inventory at the same speed as the signal goes to 0. This is also shown in \Cref{p_mass_b_TR} in comparison with \Cref{p_mass_b}: the probability that the signal is at zero when the inventory is empty is roughly 0.3 in the former case and 0.5 in the latter.
\vspace{-0.5cm}

\begin{figure}[!ht]\centering
\includegraphics[width=0.65\textwidth]{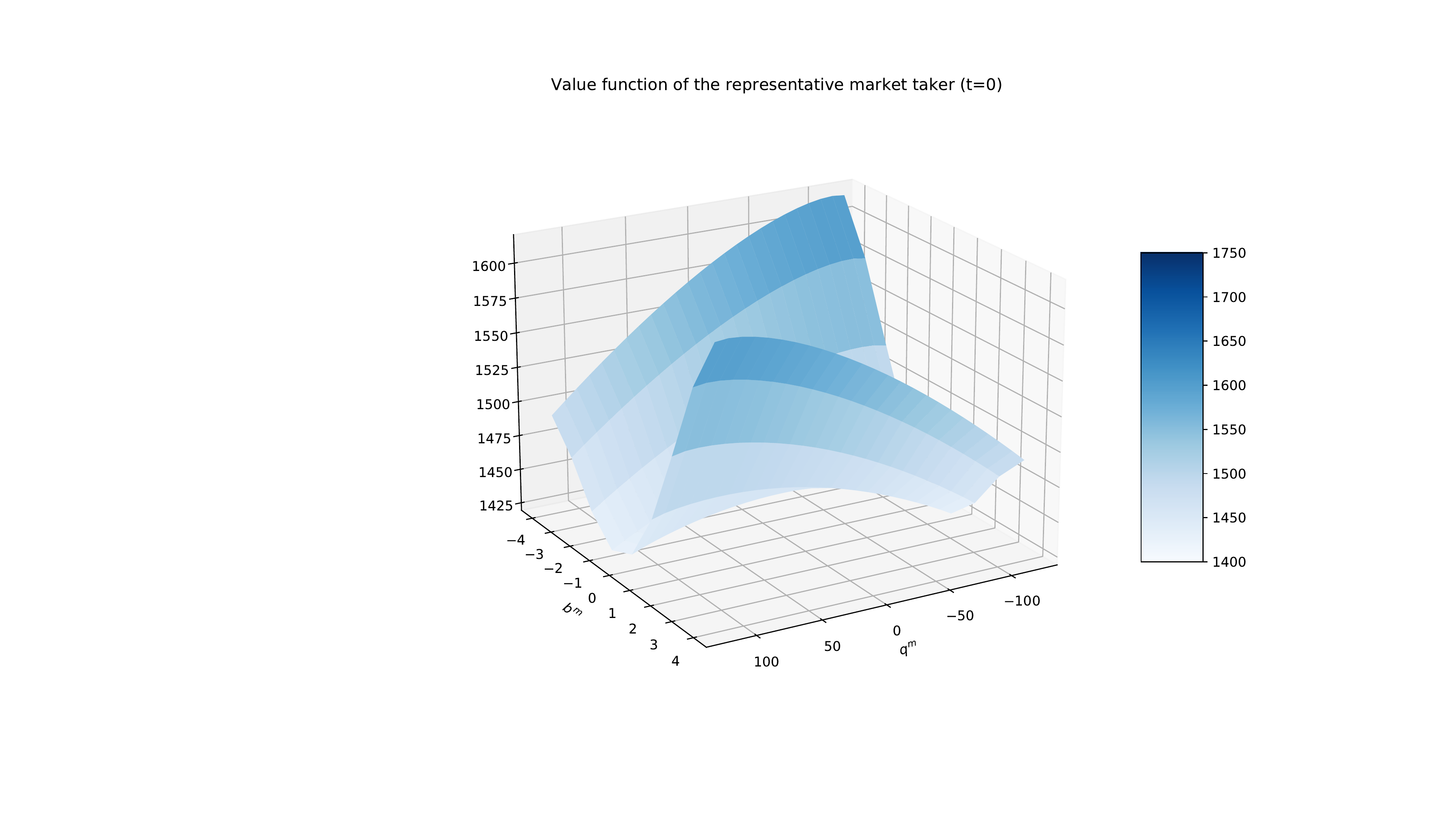}

\vspace{-2em}
\caption{\small Value function of the representative market-taker.}\label{V_minor_3D_TR}
\end{figure}

\begin{figure}[!ht]
\centering
\hspace*{-2.5cm}
\begin{subfigure}{.8\textwidth}
  \centering
  \includegraphics[width=.7\linewidth]{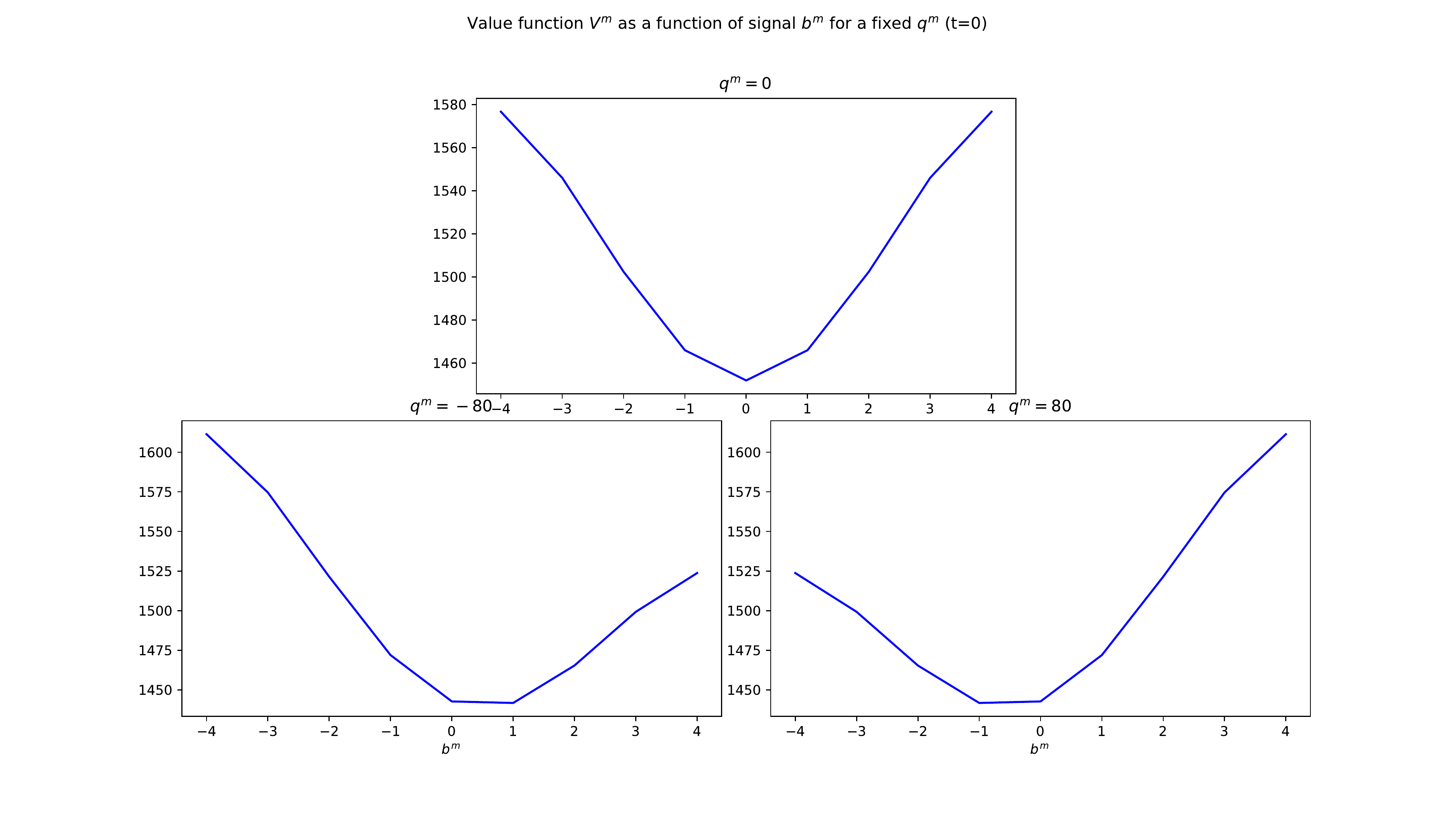}
  \vspace{-2em}
  \caption{\small For a fixed $q^m$.}
  \label{V_minor_b_TR}
\end{subfigure}%
\hspace*{-6cm}
\begin{subfigure}{.8\textwidth}
  \centering
  \includegraphics[width=.7\linewidth]{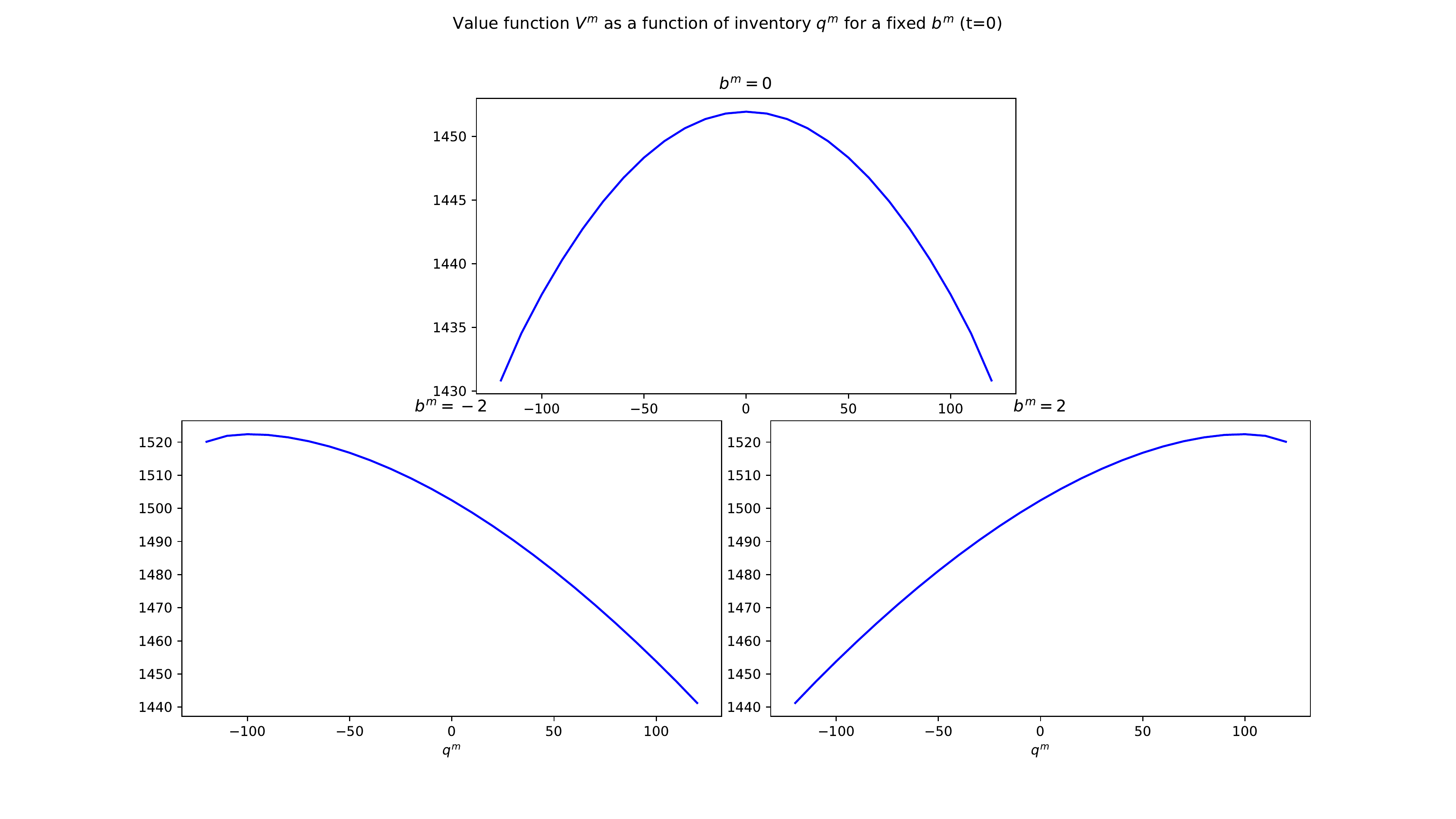}
  \vspace{-2em}
  \caption{\small For a fixed $b^m$.}
  \label{V_minor_q_TR}
\end{subfigure}
\caption{Value function of the representative market-taker.}
\label{V_minor_slices_TR}
\end{figure}

We now plot as above in \Cref{V_minor_3D_TR}, \Cref{V_minor_b_TR}, and \Cref{V_minor_q_TR} the value function $W^m$ of the market-taker at time $0$ as a function of $q^m$ and $b^m$. Whereas \Cref{V_minor_3D_TR} and \Cref{V_minor_q_TR} do not change much in aspect in comparison to \Cref{V_minor_3D} and \Cref{V_minor_q}, except that \Cref{V_minor_3D_TR} shows less difference between its highest and lowest level, due to the rapid change of the signal. \Cref{V_minor_b_TR} is rather interesting because, unlike in \Cref{V_minor_b} the minimum of the value function when $q^m = -80$ (resp. $q^m = 80$) is now reached at $1$ (resp. $-1$) instead of $0$. Unlike in the previous case, it is now less certain that the market-taker will be able to benefit from a slightly positive signal when he is short, because he cannot trade fast enough.

\vspace{-2em}

\begin{figure}[!ht]\centering
\includegraphics[width=0.7\textwidth]{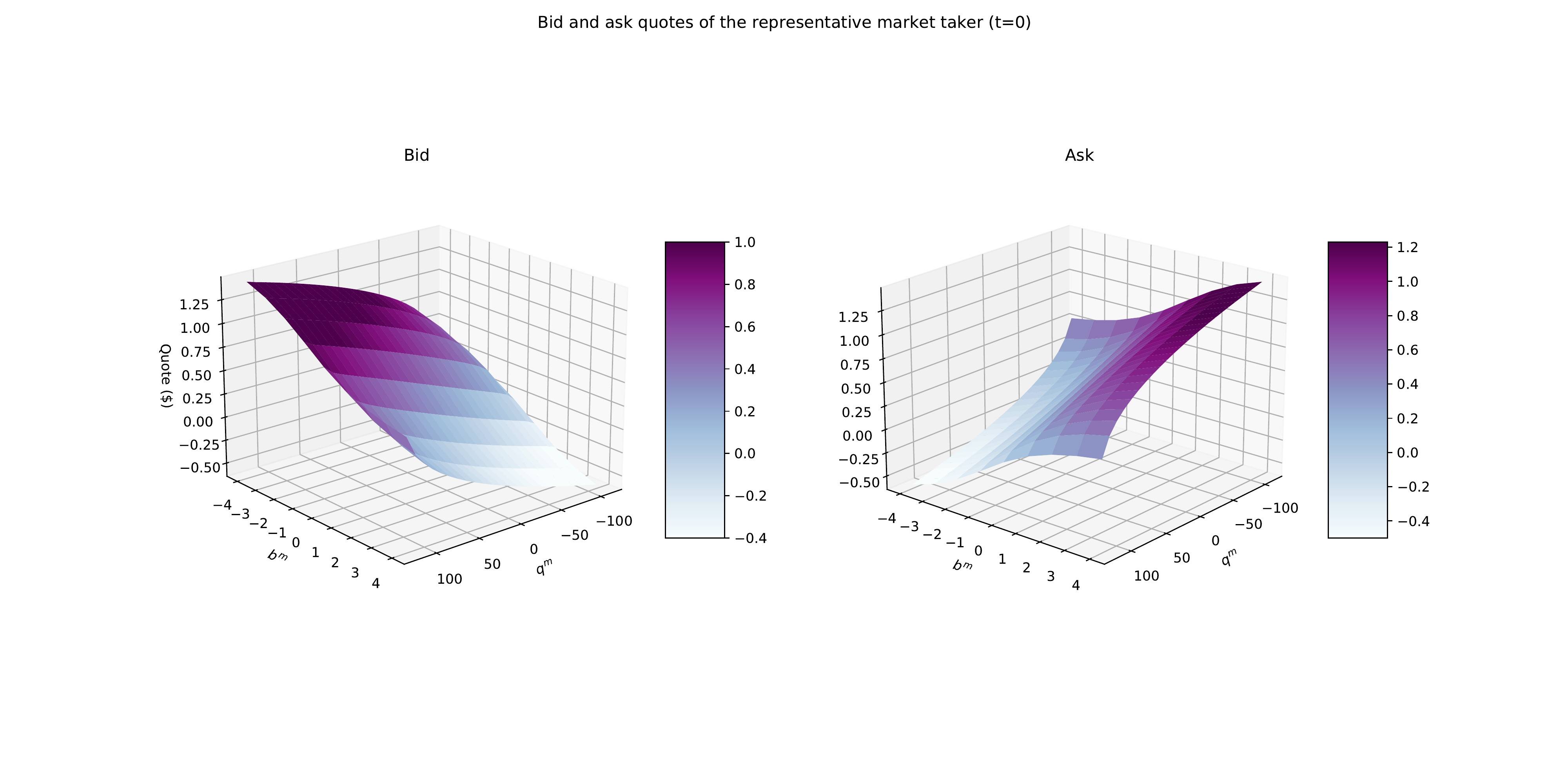}

\vspace{-4em}
\caption{\small Optimal bid and ask quotes of the representative market-taker.}\label{Quotes_minor_3D_TR}
\end{figure}

\begin{figure}[!ht]
\centering
\hspace*{-2.5cm}
\begin{subfigure}{.8\textwidth}
  \centering
  \includegraphics[width=.7\linewidth]{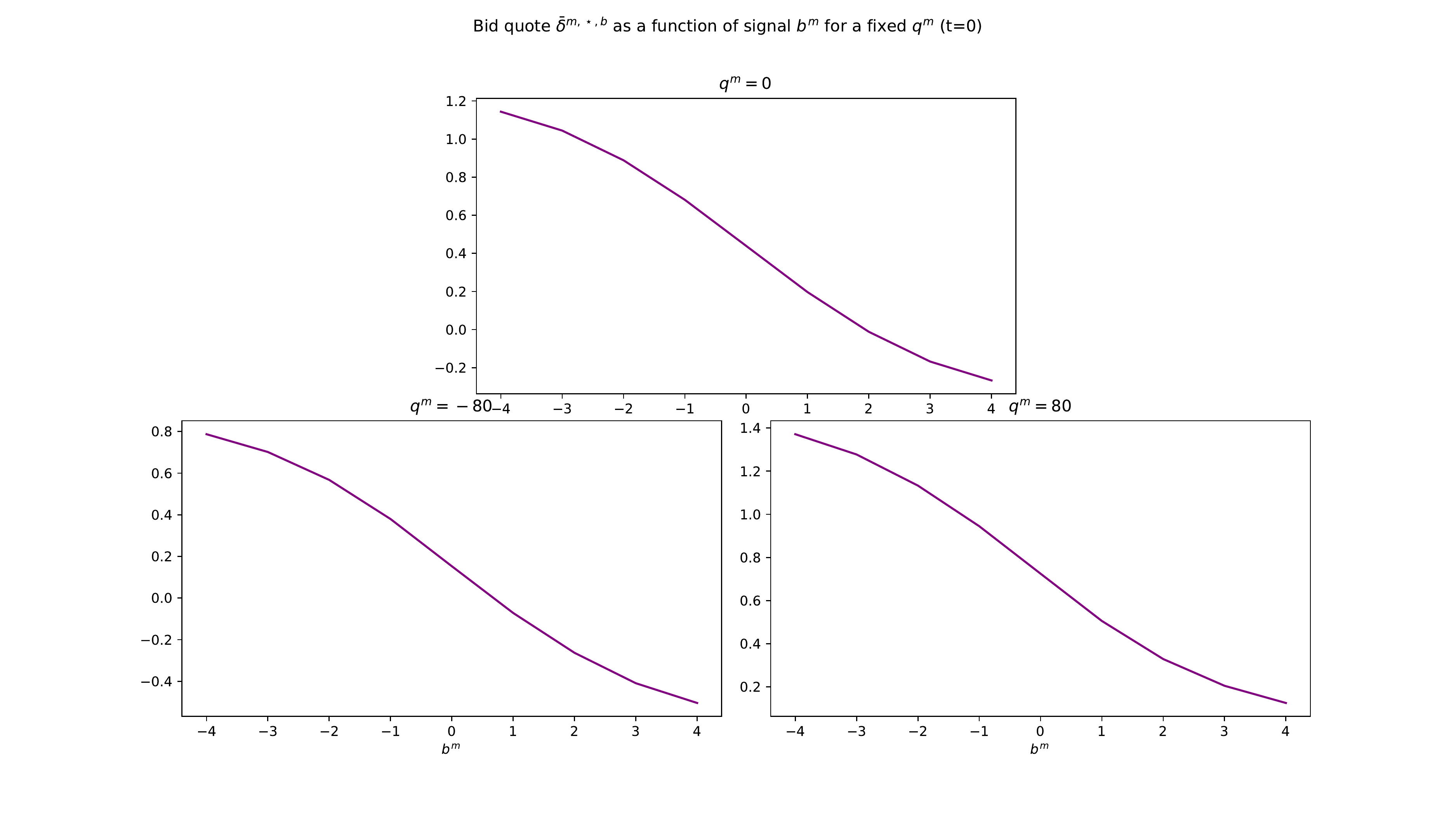}
  \vspace{-2em}
  \caption{\small For a fixed $q^m$.}
  \label{Bid_minor_b_TR}
\end{subfigure}%
\hspace*{-6cm}
\begin{subfigure}{.8\textwidth}
  \centering
  \includegraphics[width=.7\linewidth]{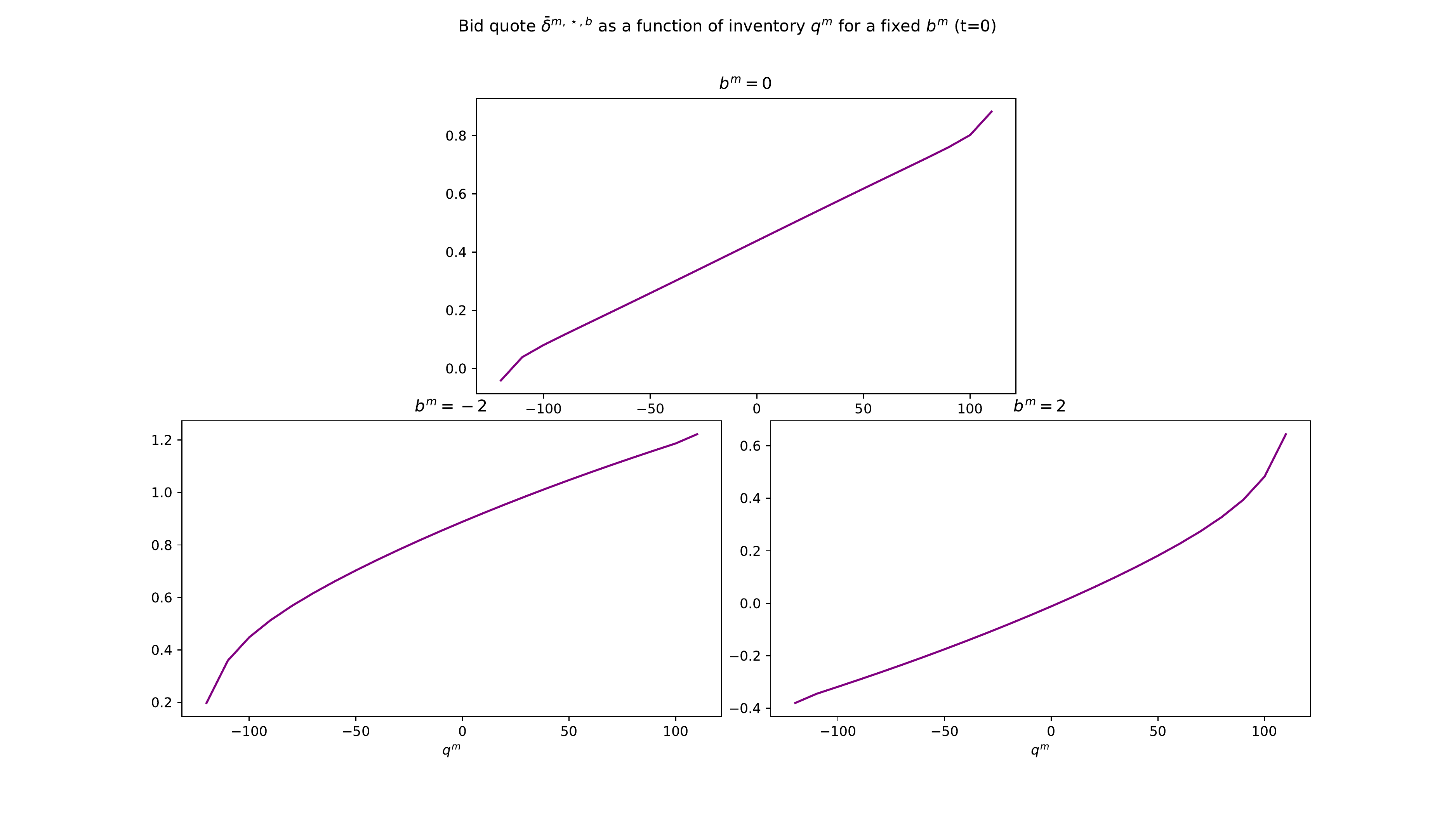}
  \vspace{-2em}
  \caption{\small For a fixed $b^m$.}
  \label{Bid_minor_q_TR}
\end{subfigure}
\caption{Optimal bid quote of the representative market-taker.}
\label{Bid_minor_TR}
\end{figure}

We then plot in \Cref{Quotes_minor_3D_TR}, \Cref{Bid_minor_b_TR}, and \Cref{Bid_minor_q_TR} the optimal quotes of the representative market-taker, which are roughly at the same level as before, except for intermediate values at which the market-taker quotes more conservatively.

\begin{figure}[!ht]\centering
\includegraphics[width=0.65\textwidth]{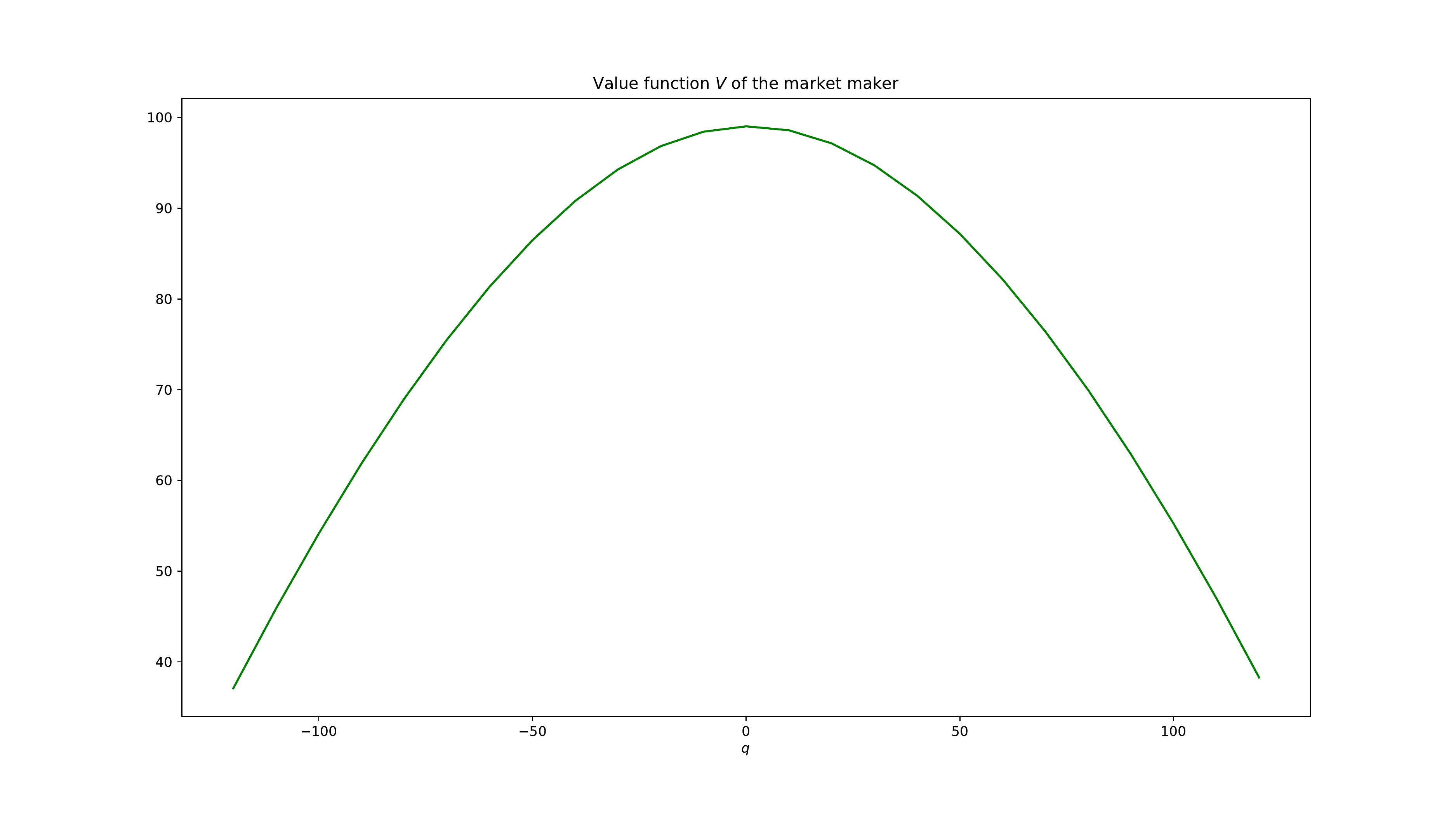}

\vspace{-2em}
\caption{\small Value function of the market-maker.}\label{V_major_TR}
\end{figure}

\begin{figure}[!ht]\centering
\includegraphics[width=0.6\textwidth]{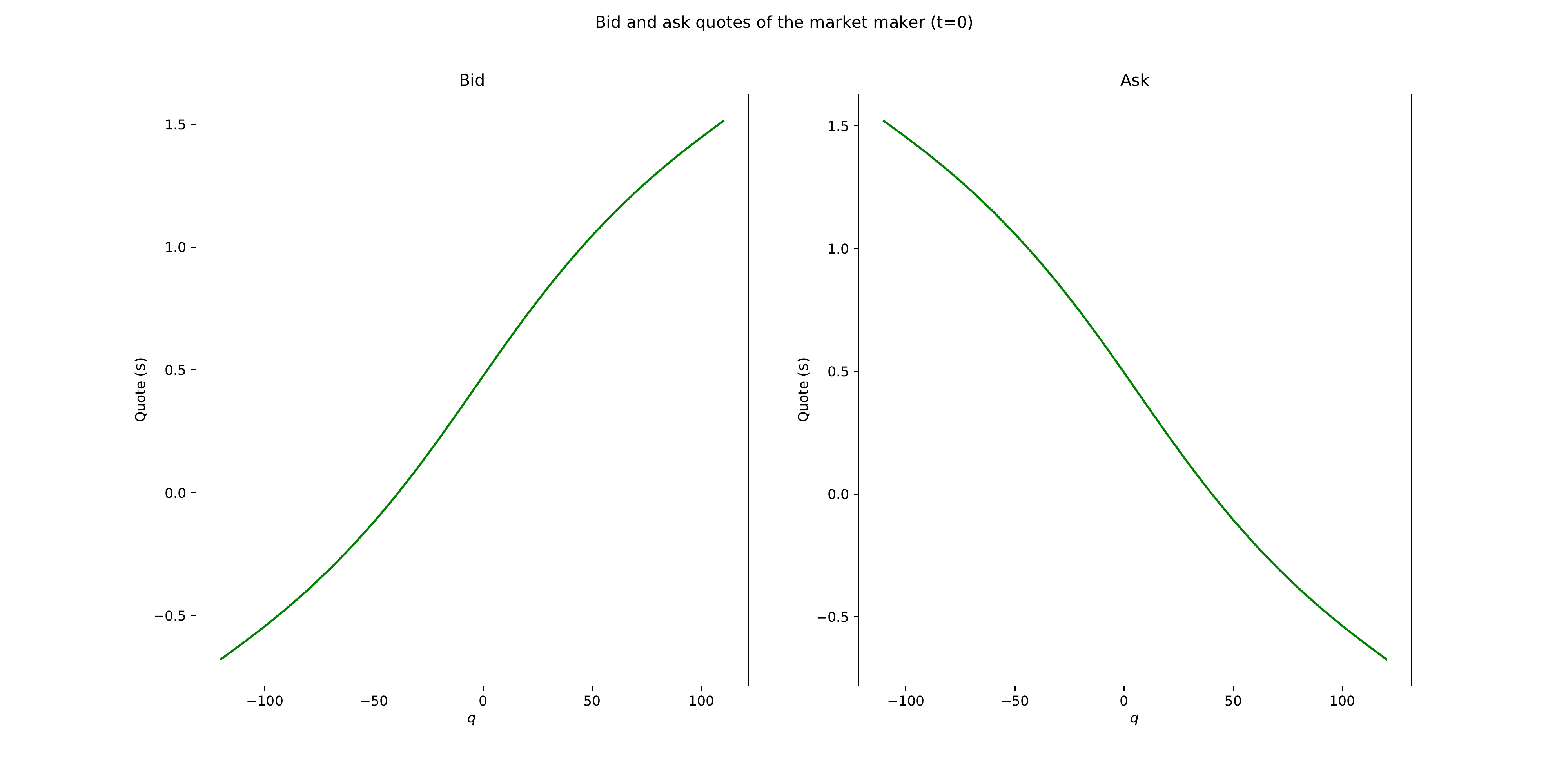}

\vspace{-2em}
\caption{\small Optimal bid and ask quotes of the market-maker.}\label{Quotes_major_TR}
\end{figure}

We finally plot in \Cref{V_major_TR} the value function at time $0$ of the market-maker, and in \Cref{Quotes_major_TR} her optimal bid and ask quotes. While the general shape of the value function and the optimal quotes do not change much when compared to \Cref{V_major} and \Cref{Quotes_major}, we see that the level of the value function is significantly lower. This is due to the fact that the market-maker trades less, because market-takers are more conservatives due to the uncertainty on the signal.

\section{Conclusion}

In this article, we tackled the problem of a market-maker on a single underlying asset facing strategical market-takers. The market-maker choses her quotes based on the behaviour of the mean-field of market-takers whose strategy is determined through an exogenous trading signal. We derive the system of HJB--Fokker--Planck equations driving this optimization problem and show existence of a solution through a fixed point argument. As the HJB equations of the market-maker and the mean-field of market-takers are decoupled, numerical resolution is simplified.\\ 

We illustrate the results of our model using a set of realistic market parameters and draw some conclusions. As expected, the probability distribution of the market-takers' inventory shifts with the exogenous trading signal and the strategy of the market-makers are adjusted with respect to this flow of probability. Finally, a more volatile exogenous signal leads to a change in inventory distribution for the market-takers: as they cannot react fast enough to a change of signal, they are not able to get back to an empty inventory at the same speed when the signal goes to zero.\\

This framework can be enhanced by introducing more complex trading signals and deeper interactions between market-maker and market-takers, at the cost of an increasing numerical complexity. In particular, one could add a common noise into the signals dynamics, and make the intensity function $\Lambda^m$ of the market-takers dependent on the behaviour of the market-maker,  but the problem then becomes numerically intractable, and theoretically much more complex to study.

\begin{appendix}

\section{Proof of Theorem \ref{existence}}\label{sec:existence}

We divide this proof in 4 steps. First we introduce suitable functional spaces to study these equations. Then we prove in the second step that, given a family $(\Pi_t)_{t\in [0,T]}$ of probability measures on $\mathcal Q^m\times \mathcal B$, there exists a unique solution to the first two equations in \eqref{MasterEq}. Conversely, in step 3, we prove that given two functions $W:[0,T]\times \mathcal Q \longrightarrow \mathbb R$ and $W^m:[0,T]\times \mathcal Q^m \times \mathcal B \longrightarrow \mathbb R$, there exists a unique solution to the last equation in \eqref{MasterEq}. Finally, we use a fixed-point theorem to conclude.\medskip

\textit{Step 1: Functional spaces}\medskip

Let us introduce the set $B(\mathcal Q^m \times \mathcal B)$ of functions $v:\mathcal Q^m \times \mathcal B \longrightarrow \mathbb R$ that are bounded on $\mathcal Q^m\times \mathcal B$. We consider the norm $\|\cdot\|_\infty$ on $B(\mathcal Q^m\times \mathcal B)$ such that for all $v \in B(\mathcal Q^m\times \mathcal B),$ 
\[
\|v\|_\infty = \underset{(q^m,b^m)\in \mathcal Q^m\times \mathcal B}{\sup} |v(q^m,b^m)|.
\]
 Then $(B(\mathcal Q^m\times \mathcal B), \|\cdot\|_\infty )$ is a Banach space. We define similarly $B(\mathcal Q)$. We will also denote by $\bar B(0,1)$ the closed ball of center $0$ and radius $1$ in $B(\mathcal Q^m\times \mathcal B)$. Note that, as $\mathcal Q^m\times \mathcal B$ is a finite set, $B(\mathcal Q^m\times \mathcal B)$ has finite dimension and $\bar B(0, 1)$ is compact.\medskip 

\textit{Step 2: HJB equations}\medskip

If we consider a continuous function $\tilde p  : t \in [0,T] \longmapsto  \tilde p_t \in \bar B(0,1)$ such that $\tilde p_t$ is a probability mass function on $\mathcal Q^m \times \mathcal B$, and if we denote by $\tilde \Pi$ the probability flow associated with $\tilde p$, we can then define $\Wc = \big( \tilde W, \tilde W^m \big)$ solution to the differential equation 
\begin{align}
\label{edoW}
\partial_t \Wc (t) = F\big(\tilde p_t,\tilde  W(t)\big),\;
\Wc(T)(q, q^m, b^m) = (0,0) \; \quad \forall (q, q^m, b^m) \in \mathcal Q \times \mathcal Q^m \times \mathcal B, 
\end{align}
where the unknown is a function $\Wc : [0,T] \longrightarrow B( \mathcal Q) \times B( \mathcal Q^m \times \mathcal B)$,  and we denote for all $(q, q^m, b^m) \in \mathcal Q \times \mathcal Q^m \times \mathcal B$:
\[\Wc(t)(q,q^m,b^m) = \big( \tilde W(t)(q), \tilde W^m(t)(q^m,b^m) \big).\]
The function $F:(p, \Wc) \in  \bar B(0,1) \times B( \mathcal Q) \times B( \mathcal Q^m \times \mathcal B) \longmapsto F(p,\Wc) \in B( \mathcal Q) \times B( \mathcal Q^m \times \mathcal B)$ is defined by the first two equations in \Cref{MasterEq}, \emph{i.e.}
\begin{align*}
   & F\big(p,(W,W^m)\big) (q,q^m,b^m)\\
   &= \bigg(\phi \sigma^2  q^2- q \kappa \int_{\mathbb R^2}   \Lc \Big(y, \bar{d}^{m}\big(W^m(t,z,y)- W^m(t,z,y+1) \big),  \bar{d}^{m}\big(W^m(t,z,y)- W^m(t,z,y-1) \big) \Big) p(t,y,z)\mathrm{d}y \d z \\
    &\quad  - \mathcal{H}\Big(t,q,\bar{d}\big(W(t,q)- W(t,q+1) \big), \bar{d}\big( W(t,q), W(t,q-1) \big),W(t,q),W(t,q+1),W(t,q-1), \bar \delta^m, p(t,\cdot,\cdot)\Big), \\
    &\quad +\gamma \sigma^2 (q^m)^2 + b^mq^m + \Kc\Big(b^m,W^m (t,q^m,b^m),W^m(t,q^m,b^m+1),W^m (t,q^m,b^m-1)\Big)\\
   &\quad- q^m \kappa \int_{\mathbb R^2}    \Lc \Big(y, \bar{d}^{m}\big(W^m(t,z,y)- W^m(t,z,y+1) \big),  \bar{d}^{m}\big(W^m(t,z,y)- W^m(t,z,y-1) \big) \Big) p(t,y,z)\mathrm{d}y \d z \\
   &\quad - \mathcal{H}^m \Big(q^m,\bar{d}^{m}\big(W^m(t,q^m,b^m)- W^m(t,q^m+1,b^m) \big),\bar{d}^{m}\big( W^m(t,q^m,b^m), W^m(t,q^m-1,b^m) \big),\\
   & \qquad \qquad W^m(t,q^m,b^m), W^m(t,q^m+1,b^m),W^m(t,q^m-1,b^m)\Big) \bigg),
\end{align*}
is continuous in $(p, W)$, and is Lipschitz-continuous in $W$ because the controls are bounded (this is proved in details in \citeauthor*{bergault2021size} \cite[Proposition 2]{bergault2021size}). Hence we know by Cauchy--Lipschitz's theorem that for any continuous $\tilde p : t\in [0,T] \longmapsto \tilde p_t \in \bar B(0,1)$, there exists a unique global solution $\tilde W : [0,T] \longrightarrow B( \mathcal Q) \times B(  \mathcal Q^m \times \mathcal B)$ to \eqref{edoW}.\medskip
 
\textit{Step 3: Fokker--Planck equation}\medskip

 For a given pair of functions $\Wc = (\tilde W, \tilde W^m)$, we can write the Fokker--Planck equation in \eqref{MasterEq} as 
 \begin{equation}
 \label{edomu}
\begin{cases}
 \partial_t p(t,\cdot,\cdot) = G(\tilde W^m(t), p(t,\cdot,\cdot)),\\
 p(0,\cdot,\cdot) = P_0,
\end{cases}     
 \end{equation}
 where the unknown is a function $p : [0,T] \longrightarrow B(\mathcal Q^m\times \mathcal B).$ The function 
 \[
 G : (W^m, p) \in  B(  \mathcal Q^m \times \mathcal B)\times B(  \mathcal Q^m \times \mathcal B) \longmapsto G(W,p) \in B(  \mathcal Q^m \times \mathcal B),
 \]
 
defined by the last equation in \eqref{MasterEq}, \emph{i.e.}
\begin{align*}
    G\big(W^m, p\big)  (q^m,b^m)&=  \mathds{1}_{\{ q^{m}>- \tilde q^m\}} \Lambda^m  \Big( \bar{d}^{m}\big(W^m(t, q^m - 1,b^m)- W^m(t,q^m,b^m) \big)\Big) p(t,q^{m}-1,b^m)\\
&\quad + \mathds{1}_{\{ q^{m} < \tilde q^m\}} \Lambda^m \Big(\bar{d}^{m}\big(W^m_(t,q^m+1,b^m)- W^m(t,q^m,b^m) \big) \Big) p(t,q^{m}+1,b^m)\\
&\quad + \bar \lambda^+\mathds{1}_{\{ b^m>- \tilde b\}}  p(t,q^{m},b^m-1) -\bar \lambda^- \mathds{1}_{\{ b^m< \tilde b\}} p(t,q^{m},b^m+1)\\
&\quad- p(t,q^{m},b^m)\bigg(\mathds{1}_{\{ q^{m} < \tilde q^m\}} \Lambda^m  \Big(\bar{d}^{m}\big(W^m(t, q^m,b^m )- W^m(t,q^m+1,b^m) \big)  \Big) \\
&\quad + \mathds{1}_{\{ q^{m} >- \tilde q^m\}}\Lambda ^m \Big(\bar{d}^{m}\big(W^m(t, q^m,b^m )- W^m(t,q^m-1,b^m) \big) \Big) + \bar \lambda^- \mathds{1}_{\{ b^m>- \tilde b\}}   +\bar \lambda^+ \mathds{1}_{\{ b^m< \tilde b\}}   \bigg) ,
\end{align*}
is continuous in $(W^m, p)$ and Lipschitz-continuous (even linear) in $p.$ Hence we know again from Cauchy--Lipschitz's theorem that for any continuous $\tilde W^m: t \in [0,T] \longrightarrow \tilde W^m(t) \in B(\mathcal Q^m\times \mathcal B)$, there exists a unique global solution $p : [0,T] \longrightarrow B(\mathcal Q^m\times \mathcal B)$ to \Cref{edomu}. Moreover, as \eqref{edomu} is actually a Fokker--Planck equation, we know that the solution  $p$ verifies $p(t,\cdot,\cdot) \in \bar B(0,1)\ \forall t \in [0,T].$\medskip 

\textit{Step 4: fixed-point theorem}\medskip

Finally, let us define $X$ the set of time-dependent probability measures $\tilde p \in \mathcal C^0([0,T], \bar B(0,1))$ such that 
\begin{equation}
\label{conditMu}
\| \tilde p(t,\cdot,\cdot)  - \tilde p(s,\cdot,\cdot) \|_\infty \leq L|t-s|,  \; \forall \ s,t \in [0,T],
\end{equation}
where $ L= 4\big(  \Lambda ^{m}( - \delta_\infty) +  \bar \lambda \big).$ Note that, by Arzel\`a--Ascoli theorem, it is clear that $X$ is relatively compact for the uniform distance. If we take $\tilde p \in X$, we can associate to it the function $\tilde W$ solution to \eqref{edoW}, and from the associated control we can get a new probability mass function $p$ solution to \eqref{edomu}. Moreover, by definition of $L$ and of the function $G$ in \eqref{edomu}, it is clear that $p$ satisfies \eqref{conditMu}, hence $p \in X$. This defines a map $\phi : \tilde p \in X \longrightarrow p= \phi(\tilde p)\in X.$

\medskip
Let us now consider $(\tilde p ^\ell)_{\ell\in\N}$ a sequence in $X$ converging uniformly to $\tilde p \in X.$ For each $\ell\in\N$, let $\Wc^\ell$ and $p^\ell$ be the corresponding solutions to \eqref{edoW} and \eqref{edomu}, respectively. By continuity of $F$ and $G$, $(\Wc^\ell)_{\ell\in\N}$ will converge to the unique solution $\Wc$ to the HJB equation \eqref{edoW} associated with $\tilde p,$ and in the same way $(p^\ell)_{\ell\in\N}$ converges to the unique solution $p$ to the Fokker--Planck equation \eqref{edomu} associated with $\Wc$. This shows the continuity of $\phi$.

\medskip
By Schauder's fixed-point theorem, we conclude that $\phi$ has a fixed-point $p$, and by denoting $\Wc = (W, W^m)$ the associated solution to \eqref{edoW}, we get that $(W, W^m, p)$ is solution to the master equation \eqref{MasterEq}.

\section{Proof of Theorem \ref{verif}}\label{sec:verif}
\begingroup
\allowdisplaybreaks
We fix $t \in [0,T)$.

\medskip
\textit{Step 1:} First, let us look at the problem of the market-maker. We introduce the arbitrary controls $(\delta^{b}_s)_{s \in [t,T]}$ and $(\delta^{a}_s)_{s \in [t,T]}$ for the market-maker, and denote by $(q_s)_{s \in [t,T]}$ her inventory process starting at time $t$, respectively at the value $q \in \mathcal Q$, and controlled by $(\delta^{b}_s)_{s \in [t,T]}$ and $(\delta^{a}_s)_{s \in [t,T]}$. Then, under $\mathbb P^{\delta, \bar  \delta^{m,\star}, \mathcal V (\bar  \delta^{m,\star},  \tilde \Pi)}$, we have
\begin{align*}
   \tilde W(T, q_T) &= \tilde W(t, q) + \int_t^T \partial_t \tilde W(s, q_s) \d s + \int_t^T \big(\tilde W(s, q_{s-}+1) - \tilde W(s, q_{s-}) \big) \d  N^{b}_s  + \int_t^T \big(\tilde W(s, q_{s-}-1) - \tilde W(s, q_{s-}) \big) \d N^{a}_s\\
   & =  \tilde W(t, q) + \int_t^T \partial_t \tilde W(s, q_s) \d s\\
   & \quad + \int_t^T \big(\tilde W(s, q_{s-}+1) - \tilde W(s, q_{s-}) \big) \mathds{1}_{\{ q_{s-} < \tilde q\}}  \Lambda\bigg(\int_{\mathbb R^2} \delta^{m,\star,a}(s,z,y,\tilde \Pi_s)\tilde{\Pi}_s(\d y,\d z) , \delta^{b}_s\bigg) \d s\\
   & \quad + \int_t^T \big(\tilde W(s, q_{s-}-1) - \tilde W(s, q_{s-}) \big) \mathds{1}_{\{ q_{s-} >- \tilde q\}}  \Lambda\bigg(\int_{\mathbb R^2}\delta^{m,\star,b}(s,z,y,\tilde \Pi_s)\tilde{\Pi}_s(\d y,\d z) , \delta^{a}_s\bigg) \d s\\
   &  \quad  + \int_t^T \big(\tilde W(s, q_{s-}+1) - \tilde W(s, q_{s-}) \big) \d \tilde N^{b}_s + \int_t^T \big(\tilde W(s, q_{s-}-1) - \tilde W(s, q_{s-}) \big)\d \tilde N^{a}_s,
\end{align*}
where $\tilde N^{b}$ and $\tilde N^{a}$ are the compensated processes associated to $N^{b}$ and $N^{a}$ under $\mathbb P^{\delta, \bar  \delta^{m,\star}, \mathcal V (\bar  \delta^{m,\star},  \tilde \Pi)}$. We can then write
\begin{align*}
     \tilde W(T, q_T) & = \tilde W(t, q) + \int_t^T \Bigg( \partial_t \tilde W(s, q_s) - \phi \sigma^2(q_s)^2\\
     &\quad + \Big(\delta^{b}_s  + \tilde W(s, q_{s-}+1) - \tilde W(s, q_{s-})  \Big) \mathds{1}_{\{ q_{s-} < \tilde q\}}  \Lambda\bigg(\int_{\mathbb R^2} \delta^{m,\star,a}(s,z,y,\tilde \Pi_s)\tilde{\Pi}_s(\d y,\d z) , \delta^{b}_s\bigg) \\
     & \quad + \Big(\delta^{a}_s  + \tilde W(s, q_{s-}-1) - \tilde W(s, q_{s-})  \Big) \mathds{1}_{\{ q_{s-} >- \tilde q\}} \Lambda\bigg(\int_{\mathbb R^2}\delta^{m,\star,b}(s,z,y,\tilde \Pi_s)\tilde{\Pi}_s(\d y,\d z) , \delta^{a}_s\bigg)\\
     &\quad +  q_s \kappa \int_{\mathbb R^2}   \Lc \Big(y, \bar{d}^{m}\big(\tilde W^m(t,z,y)- \tilde W^m(t,z,y+1) \big),  \bar{d}^{m}\big(\tilde W^m(t,z,y)- \tilde W^m(t,z,y-1) \big) \Big) \Pi_t(\mathrm{d}y, \d z) \Bigg) \d s\\
     &\quad + \int_t^T \Bigg( \phi \sigma^2 (q_s)^2  - \delta^{b}_s  \mathds{1}_{\{ q_{s-} < \tilde q\}} \Lambda\bigg(\int_{\mathbb R^2} \delta^{m,\star,a}(s,z,y,\tilde \Pi_s)\tilde{\Pi}_s(\d y,\d z) , \delta^{b}_s\bigg)\\
     &\quad - \delta^{a}_s   \mathds{1}_{\{ q_{s-} >- \tilde q\}} \Lambda\bigg(\int_{\mathbb R^2}\delta^{m,\star,b}(s,z,y,\tilde \Pi_s)\tilde{\Pi}_s(\d y,\d z) , \delta^{a}_s\bigg)\\
    &\quad  - q_s \kappa \int_{\mathbb R^2}   \Lc \Big(y, \bar{d}^{m}\big(\tilde W^m(t,z,y)- \tilde W^m(t,z,y+1) \big),  \bar{d}^{m}\big(\tilde W^m(t,z,y)- \tilde W^m(t,z,y-1) \big) \Big) \Pi_t(\mathrm{d}y, \d z) \Bigg)\d s\\
      &  \quad  + \int_t^T \big(\tilde W(s, q_{s-}+1) - \tilde W(s, q_{s-}) \big)\d \tilde N^{b}_s + \int_t^T \big(\tilde W(s, q_{s-}-1) - \tilde W(s, q_{s-}) \big)\d \tilde N^{a}_s.
\end{align*}
From the HJB equation solved by $\tilde W$ and the associated terminal condition $\tilde W(T, .) = 0$, we deduce
\begin{align*}
    0 &\leq \tilde W(t, q) + \int_t^T \bigg( \phi \sigma^2 (q_s)^2  - \delta^{b}_s  \mathds{1}_{\{ q_{s-} < \tilde q\}} \Lambda\bigg(\int_{\mathbb R^2} \delta^{m,\star,a}(s,z,y,\tilde \Pi_s)\tilde{\Pi}_s(\d y,\d z) , \delta^{b}_s\bigg)\\
     &\quad - \delta^{a}_s   \mathds{1}_{\{ q_{s-} >- \tilde q\}} \Lambda\bigg(\int_{\mathbb R^2}\delta^{m,\star,b}(s,z,y,\tilde \Pi_s)\tilde{\Pi}_s(\d y,\d z) , \delta^{a}_s\bigg)\\
    &\quad  - q_s \kappa \int_{\mathbb R^2}   \Lc \Big(y, \bar{d}^{m}\big(\tilde W^m(t,z,y)- \tilde W^m(t,z,y+1) \big),  \bar{d}^{m}\big(\tilde W^m(t,z,y)- \tilde W^m(t,z,y-1) \big) \Big) \Pi_t(\mathrm{d}y, \d z) \bigg)\d s\\
      &  \quad  + \int_t^T \big(\tilde W(s, q_{s-}+1) - \tilde W(s, q_{s-}) \big)\d \tilde N^{b}_s+ \int_t^T \big(\tilde W(s, q_{s-}-1) - \tilde W(s, q_{s-}) \big)\d \tilde N^{a}_s.
\end{align*}
Notice that $\tilde N^{b}$ and $\tilde N^{a}$ are martingales under $\mathbb P^{\delta, \bar  \delta^{m,\star}, \mathcal V (\bar  \delta^{m,\star},  \tilde \Pi)}$, since by definition of $\mathcal A_\infty$, their intensities are bounded. Taking expectation yields
\begin{align}\label{verifMM}
\begin{split}
 \tilde W(t, q) &\geq \mathbb E^{\delta, \bar  \delta^{m,\star}, \mathcal V (\bar  \delta^{m,\star},  \tilde \Pi)} \Bigg[ \int_t^T \bigg(-\phi \sigma^2 (q_s)^2 + \delta^{b}_s \mathds{1}_{\{ q_{s-} < \tilde q\}} \Lambda\bigg(\int_{\mathbb R^2} \delta^{m,\star,a}(s,z,y,\tilde \Pi_s)\tilde{\Pi}_s(\d y,\d z) , \delta^{b}_s\bigg)\\
     & \quad + \delta^{a}_s   \mathds{1}_{\{ q_{s-} >- \tilde q\}} \Lambda\bigg(\int_{\mathbb R^2}\delta^{m,\star,b}(s,z,y,\tilde \Pi_s)\tilde{\Pi}_s(\d y,\d z) , \delta^{a}_s\bigg)\\
     &\quad + q_s \kappa \int_{\mathbb R^2}   \Lc \Big(y, \bar{d}^{m}\big(\tilde W^m(t,z,y)- \tilde W^m(t,z,y+1) \big),  \bar{d}^{m}\big(\tilde W^m(t,z,y)- \tilde W^m(t,z,y-1) \big) \Big) \Pi_t(\mathrm{d}y, \d z) \bigg)\d s\Bigg],   
\end{split}
\end{align}
with equality when the market-maker plays the controls in \eqref{cloptcont0bcor}.

\medskip
By taking the supremum in \eqref{verifMM}, we get that $\tilde W (0, q_0) =  \tilde V (\bar \delta^m, \tilde \Pi)$, and the controls given in \eqref{cloptcont0bcor} are optimal for the market-maker when the market-taker plays the controls given again in \eqref{cloptcont0bcor}.

\medskip
\textit{Step 2:} We show by the same argument that $\tilde W^m (0, q^m_0, b^m_0) =  \tilde V^m (\bar \delta^m, \tilde \Pi)$ and that the controls given in \eqref{cloptcont0bcor} are optimal for the market-taker when the market-maker plays his respective controls in \eqref{cloptcont0bcor}.
\endgroup
\end{appendix}

\small
\bibliography{mainMT.bib}
\end{document}